%% file: variationalSurrogateOpt.tex
\documentclass[3p, number,sort&compress,times,11pt,fleqn]{elsarticle}
\graphicspath{{./figs/}}
\usepackage{url}
\usepackage{booktabs}
\usepackage{multirow}
\usepackage[table]{xcolor}
\usepackage{setspace}
\usepackage{algorithm}
\usepackage{algpseudocode}
\algtext*{EndWhile}
\algtext*{EndFor}
\algtext*{EndIf}
\usepackage[colorlinks = true,
linkcolor = blue,
urlcolor  = blue,
citecolor = blue,
anchorcolor = blue]{hyperref}
\usepackage{csml}

\journal{Elsevier}

\begin{document}

\begin{frontmatter}

\title{Variational Bayesian surrogate modelling with application to robust design optimisation}

\author{Thomas A. Archbold}
\author{Ieva Kazlauskaite}
\author{Fehmi Cirak\corref{cor1}}
\ead{f.cirak@eng.cam.ac.uk}

\cortext[cor1]{Corresponding author}

\address{Department of Engineering, University of Cambridge, Trumpington Street, Cambridge, CB2 1PZ, UK }

\begin{abstract}

Surrogate models provide a quick-to-evaluate approximation to complex computational models and are essential for multi-query problems like design optimisation. The inputs of current deterministic computational models are usually high-dimensional and uncertain. We consider Bayesian inference for constructing statistical surrogates with input uncertainties and intrinsic dimensionality reduction. The surrogate is trained by fitting to data obtained from a  deterministic computational model. The assumed prior probability density of the surrogate is a Gaussian process. We determine the respective posterior probability density and parameters of the posited statistical model using variational Bayes. The non-Gaussian posterior is approximated by a Gaussian trial density with free variational parameters and the discrepancy between them is measured using the Kullback-Leibler (KL) divergence. We employ the stochastic gradient method to compute the variational parameters and other statistical model parameters by minimising the KL divergence. We demonstrate the accuracy and versatility of the proposed reduced dimension variational Gaussian process (RDVGP) surrogate on illustrative and robust structural optimisation problems where cost functions depend on a weighted sum of the mean and standard deviation of model outputs. 
 
\end{abstract}


\end{frontmatter}

\input{introduction}

\input{design_optimisation}

\input{statistical_model}

\input{examples}

\input{conclusion}

\input{appendix}

\bibliographystyle{elsarticle-num}
\bibliography{references}

\end{document}

%% file: introduction.tex
\section{Introduction}\label{section:introduction}
The design of engineering products relies on complex computational models typically consisting of several submodels responsible for different aspects of the design process, like geometric modelling, stress analysis, etc. However, these computational models are costly to evaluate and quickly become a bottleneck in multi-query problems such as optimisation, uncertainty quantification, inverse problems, and model calibration. Surrogates address this problem by providing a quick-to-evaluate approximation to costly  computational models. In surrogate modelling, the computational model is treated as a black-box function that can be accessed  by only recording its outputs for given inputs. The hyperparameters of the surrogate are learned offline by fitting to data collected from the black-box function. Common surrogate modelling techniques include polynomial chaos expansions~\cite{xiu2002modeling,soize2004physical,zhang2023stochastic,kantarakias2023sensitivity}, neural networks~\cite{nielsen2015neural,papadrakakis2002reliability,bishop2024deep,haghighat2021physics} and Gaussian process (GP) regression~\cite{williams2006gaussian,santner2003design,martin2005use,forrester2008engineering,koh2023stochastic}. Although most commonly used computational models are deterministic, it is critical to consider the input uncertainties to achieve more efficient and robust designs~\cite{oden2010computerI,roy2011comprehensive}. These uncertainties typically arise from variations in manufacturing processes and insufficiently known operating conditions. Such intrinsic aleatoric uncertainties are critical for our primary application of robust design optimisation (RDO); see e.g.~\cite{o2006bayesian,der2009aleatory} for a discussion on aleatoric and epistemic uncertainties. In RDO~\cite{schueller2008computational,doltsinis2004robust,asadpoure2011robust,da2017stress,ben2024robust}, the model inputs consist of design variables like geometric dimensions and fixed (immutable) variables like forcing or constitutive parameters, while the model outputs are quantities of interest like the maximum stress or compliance. RDO yields a design that is less sensitive to the input variations by considering a cost function consisting of the weighted sum of the expectation and standard deviation of typical objective and constraint functions used in deterministic optimisation. Surrogate-based techniques are widely used in RDO, see e.g.~\cite{sakata2003structural,zhang2017sequential,tootkaboni2012topology,bessa2018design,zafar2020efficient,keshavarzzadeh2020stress,tian2020enhanced}.

\begin{figure}[t!]
\centering
\subfloat[True probability density $p(f(\vec{s}))$]{\includegraphics[width=80mm]{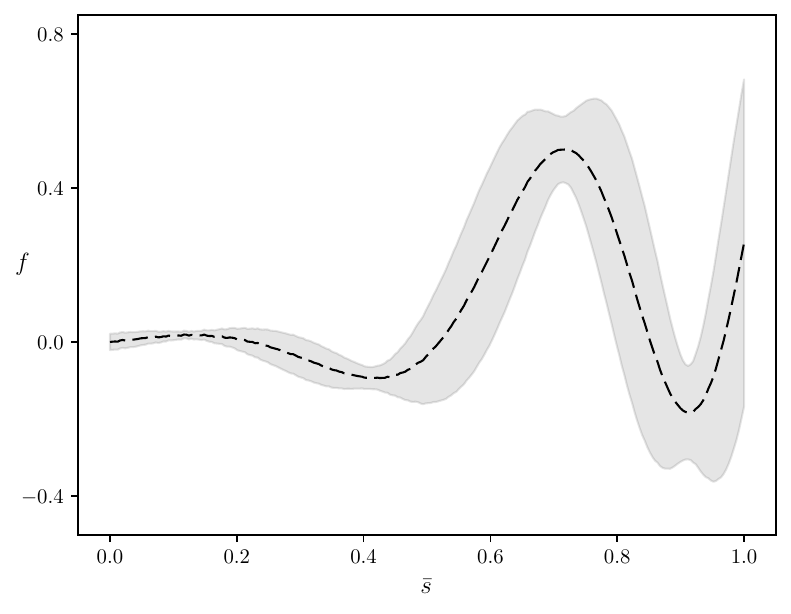}}
\centering
\subfloat[Training data points ($n=5$)]{\includegraphics[width=80mm]{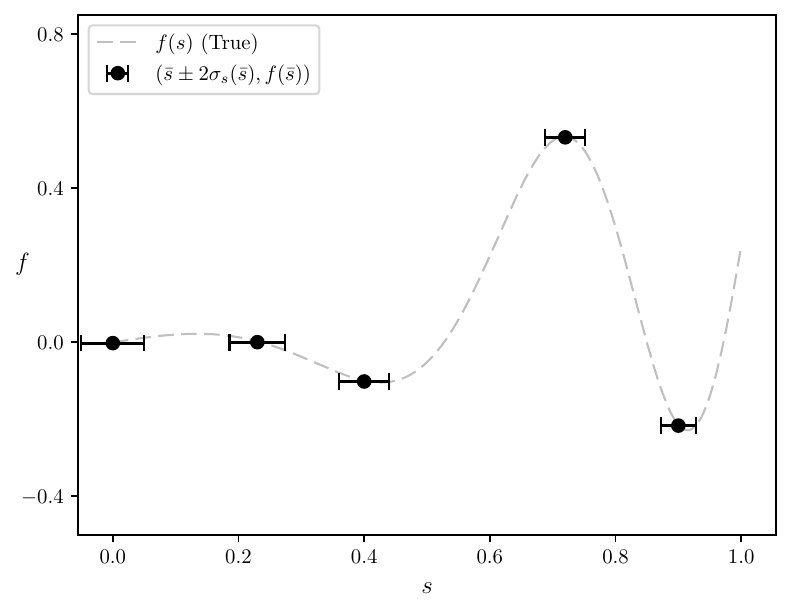}}\\
\centering
\subfloat[Posterior density of the standard GP surrogate]{\includegraphics[width=80mm]{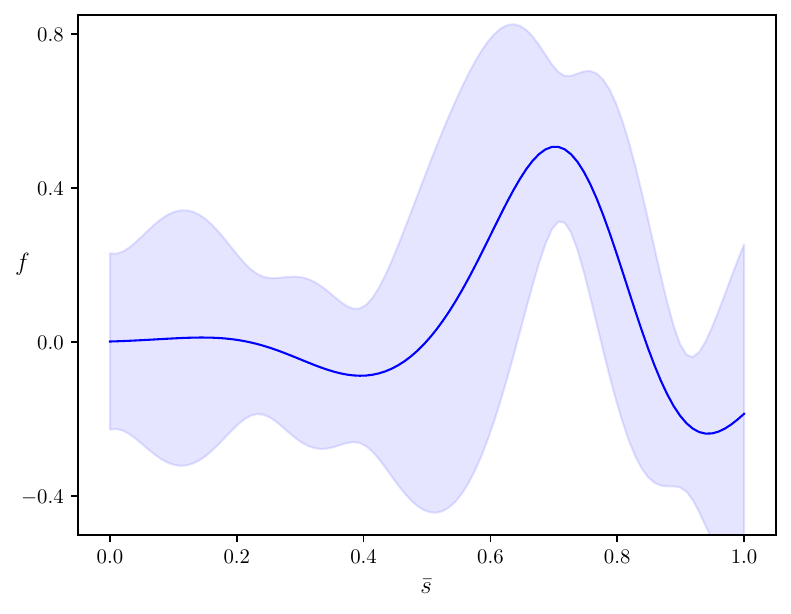}}
\centering
\subfloat[Posterior density of the proposed RDVGP surrogate]{\includegraphics[width=80mm]{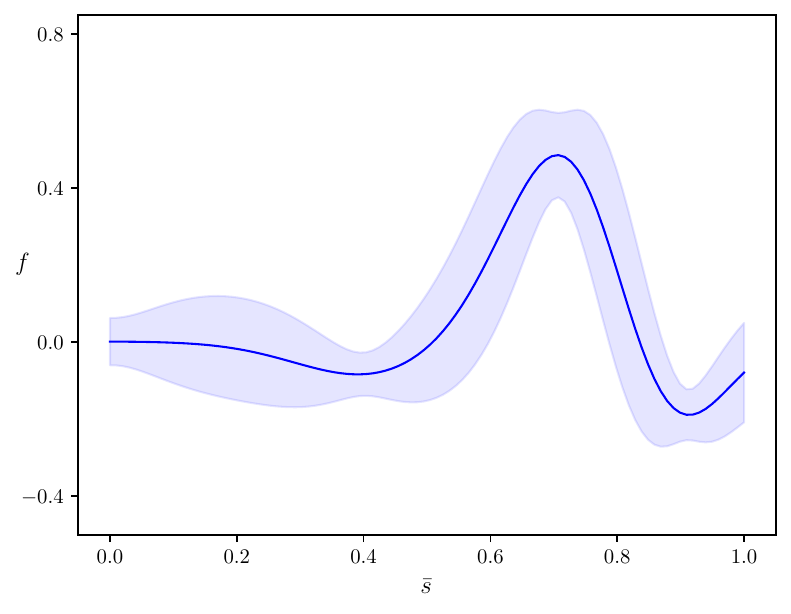}}
\caption{Comparison of standard GP and the proposed RDVGP surrogates for a function \mbox{$f(s)= -0.5s\sin{(3\pi s^2)} + 0.25s$} and the normally distributed input variable with spatially dependent variance~\mbox{$p(s) = \mathcal{N}(\bar{s},\sigma_s^2)$}; see also Section \ref{subsection:illustrative_example_1}. In the plots the lines indicate the mean and the shaded areas the~$95\%$ confidence intervals. Shown is (a) the true probability density~\mbox{$p(f(\bar{s})) = \int p(f\vert \vec{s})p(\vec{s}) \D \vec{s}$} obtained by MC sampling, (b)~the~$n=5$ training data points comprising the training set~$\set D$ and their spatially varying standard deviations, (c)~the inferred standard GP posterior probability density~$p( f(\bar{s}) \vert \set D)$, and (d) the inferred RDVGP posterior probability density~$p( f(\bar{s}) \vert \set D)$. Comparing (a), (c), and (d) it is evident that the RDVGP posterior is much closer to the true probability density than the standard GP posterior.}
\label{fig:rdvgp_comparison_intro}
\end{figure}

We construct a statistical surrogate model by fitting to data sampled from the costly-to-evaluate deterministic computational model. As common in probabilistic approaches, most prominently GP regression, we pose surrogate model construction as a Bayesian inference problem~\cite{kennedy2001bayesian,murphy2012machine,gelman2013bayesian,grigo2019bayesian,girolami2021statistical}.  That is, the posterior probability density representing a surrogate~$f(\vec s)$ is determined by updating a prior probability density in light of a training data set~$\set D$ using Bayes' rule. The prior probability density for~$f(\vec s)$ is a GP with a prescribed mean and covariance with several hyperparameters. The training data set~\mbox{$\set D= \{(\vec s_i, \,  y_i ) \}_{i=1}^{n}$} collects the~$n$ evaluations of the computational model with each pair consisting of an input~$\vec s_i \in \mathbb R^{d_s}$ and its output~$y_i \in \mathbb R$.  The assumed prior probability density defines a distribution over functions with a certain mean and covariance. The required likelihood function is obtained from the posited statistical observation model~$y=f(\vec s)+ \epsilon_y $  with an additive noise or error term~$\epsilon_y $.   A crucial difference of the sketched approach from standard GP regression is that the input vector~$\vec s$  has intrinsic aleatoric uncertainties, e.g. due to manufacturing, see Figure~\ref{fig:rdvgp_comparison_intro}. Consequently, the posterior probability density representing the surrogate~$f(\vec s)$ encodes in addition to epistemic uncertainty, also  intrinsic aleatoric uncertainties that do not vanish in the limit of infinite data. In standard GP regression with a deterministic input~$\vec s$, all uncertainty is epistemic and tends toward zero when more data is taken into account~\cite{williams2006gaussian,o2006bayesian}. The key challenge in Bayesian inference with an uncertain input~$\vec s$ is that the posterior density is not analytically tractable (even when both the prior probability density and the likelihood functions are GPs).  Hence, it is necessary to use approximation techniques, like Markov chain Monte Carlo (MCMC), Laplace approximation, or variational Bayes (VB)~\cite{gelman2013bayesian}.  

In design optimisation, as in most other multi-query applications, the dimension of the input variable~$\vec s$ must be restricted for computational tractability.  The dimension of the input variable is usually large from the outset, but generally contains a low intrinsic dimension that can be discovered using a suitable dimensionality reduction technique. Dimensionality reduction is an extensively studied topic, and many techniques are available~\cite{constantine2014active,bouhlel2016improving,gaudrie2020modeling,lam2020multifidelity,romor2021multi}. Two-step approaches for sequentially combining dimensionality reduction with standard GP regression have also been proposed~\cite{tripathy2016gaussian,guo2018reduced,tsilifis2021bayesian}. Departing from earlier approaches and adopting a fully Bayesian viewpoint, we implement dimensionality reduction by positing the statistical observation model  $y=f(\vec z) + \epsilon_y $ with the  low-dimensional (latent) input vector~$\vec z= \vec W^\trans \vec s$ and the non-square (tall) orthogonal matrix~$\vec W$, i.e. \mbox{$\vec W^\trans \vec W = \vec I$}. A reduction in input dimension is possible because in most applications the intrinsic dimensionality of the input-output manifold is low  and can be approximated as a hyperplane~\cite{van2009dimensionality}. As previously noted, the input vector~$\vec s$ and consequently, the latent vector~$\vec z$ have intrinsic aleatoric uncertainties whereas the matrix~$\vec W$ is deterministic. The posterior probability density of~$f(\vec s)$ and all statistical model parameters including the dimension of~$\vec z$ and entries of~$\vec W$, can be obtained by consequent application of the Bayes' rule. The posterior probability densities and the marginal likelihood density for determining the model parameters have no analytically tractable solution and must be approximated.  

VB provides an optimisation-based formulation for approximating the posterior density and parameters of statistical models~\cite{jordan1999introduction,blei2017variational,povala2022variational,vadeboncoeur2023fully}. In VB, the true posterior density is approximated by a trial density and the distance between the two is measured using the Kullback-Leibler (KL) divergence. The chosen trial density is a parameterised probability density with free variational parameters.  It can be shown that minimising the KL divergence is equivalent to maximising the evidence lower bound (ELBO). As the name suggests, the ELBO is a lower bound for the log-evidence, also called the log-marginal likelihood, which quantifies how well  the posited statistical observation model explains the training data. In our statistical surrogate model described above, we approximate the posterior density of the surrogate~$f(\vec s)$ and the model hyperparameters consist of the entries of the projection matrix~$\vec W$ and the hyperparameters of the prior. We select a Gaussian trial density with a diagonal covariance structure and choose its mean and variance as the variational parameters. All the model hyperparameters and the variational parameters are determined by maximising the ELBO. In the proposed implementation, we maximise the ELBO using a stochastic gradient method~\cite{hoffman2013stochastic}, specifically the Cayley ADAM optimiser~\cite{li2020efficient},  which automatically maintains the orthogonality of~$\vec W$. Querying the posterior probability density at a new set of test inputs $n_*$ with a training data set $\mathcal{D}$ of size $n$ has a computational complexity of $\mathcal{O}(n_* n^2)$. This is usually computationally too onerous for large-scale robust design optimisation. Therefore, we introduce a sparse approximation of~$f(\vec z)$ using~$m$ pseudo training data points reducing the complexity to~$\mathcal{O}(n_* m^2)$~\cite{quinonero2005unifying,titsias2009variational}. Although the number~$m$ is user chosen, the coordinates of the pseudo input points are determined as part of VB. The overall solution approach implemented is akin to the latent variable GP models widely used in machine learning~\cite{lawrence2007learning,damianou2016variational,leibfried2020tutorial}. For brevity, we refer to this as the reduced dimension variational Gaussian process (RDVGP) surrogate.

This paper is structured as follows. In Section \ref{section:robust_design_optimisation_gaussian_processes}, we define RDO problems in terms of objective and constraint functions describing the behavior of complex engineering systems, and review GP regression for emulating these functions. In Section \ref{section:statistical_model}, we present the statistical surrogate model by postulating a graphical model with intrinsic statistical assumptions and deriving the ELBO from the joint probability density using VB, later extending to include sparsity. Subsequently, we detail the algorithm used to train the RDVGP surrogate, and provide the metrics used to verify its accuracy when compared to the true solution estimated using Monte Carlo (MC) sampling. Four examples are introduced in Section \ref{section:examples} to demonstrate the accuracy and versatility of the predicted marginal posterior probability densities for solving RDO problems. Finally, Section \ref{section:conclusion} concludes the paper and provides promising directions for further research.

%% file: design_optimisation.tex
\section{Robust design optimisation and Gaussian processes}\label{section:robust_design_optimisation_gaussian_processes}

In this section, we provide the necessary background on RDO and GP regression. The objective and constraint functions in robust design optimisation depend on inputs with prescribed probability densities. In standard GP regression, the inputs are deterministic, and the posterior has a closed form. We show how the posterior becomes analytically intractable when the inputs are random. 

\subsection{Review of robust design optimisation}\label{subsection:robust_design_optimisation}

In many applications, the computational model output is not directly the quantity of interest, but an argument provided to an objective or constraint function representing model behavior, and may be computed from a finite element (FE) model. For example, in structural mechanics, compliance objectives and stress constraints are both functions of the partial differential equation (PDE) solution. For brevity, the objective $J: \mathbb{R}^{d_s} \rightarrow \mathbb{R}$ and constraint $H^{(j)}: \mathbb{R}^{d_s} \rightarrow \mathbb{R}$ may be denoted as functions of the input variables $\vec{s} \in \mathbb{R}^{d_s}$. A total of $d_y$ functions, consisting (without loss of generality) of a single objective and $(d_y-1)$ constraints are considered.

We assume the input variable probability density is prescribed a priori, which could be Gaussian \mbox{$p(\vec{s}) = \mathcal{N}(\bar{\vec{s}},\vec{\Sigma}_s)$}, where \mbox{$\bar{\vec{s}} \in \mathbb{R}^{d_s}$} is the mean and \mbox{$\vec{\Sigma}_s \in \mathbb{R}^{d_s \times d_s}$} is the covariance. For instance, the CAD embodiment of an engineering product may consist of components with geometric variation represented by a Gaussian probability density, where an engineer models the product in CAD using the mean design variables~$\bar{\vec{s}}$. Since the input variables are random, the objective $J(\vec{s})$ and constraints $H^{(j)}(\vec{s})$ are also random and the RDO problem may be solved using the weighted sum cost function
\begin{equation}\label{eq:rdo1}
\begin{aligned}
\min_{\bar{s} \in D_{\bar{s}}} \quad & \frac{(1-\alpha)}{\bar{\mu}}\expect\left(J(\vec{s})\right) + \frac{\alpha}{\bar{\sigma}}\sqrt{\var\left(J(\vec{s})\right)},\\
\textrm{s.t.} \quad & \expect\left(H^{(j)}(\vec{s})\right) + \beta^{(j)} \sqrt{\var\left(H^{(j)}(\vec{s})\right)} \leq 0, & \quad j&\in\{1,2,...,d_y-1\},\\
  & \sqrt{\var \left(H^{(j)}(\vec{s})\right)} \leq \sigma^{(j)},\\
  & D_{\bar{s}} = \left\{\bar{\vec{s}} \in \mathbb{R}^{d_s} \, \big\vert \, \bar{\vec{s}}^{(l)} \leq \bar{\vec{s}} \leq \bar{\vec{s}}^{(u)}\right\},
\end{aligned}
\end{equation}
where $\alpha \in \mathbb{R}^+$ is a prescribed weighting factor ($0 \leq \alpha \leq 1$), $\bar{\mu} \in \mathbb{R}$ and $\bar{\sigma}\in \mathbb{R}$ are normalisation constants, $\beta^{(j)} \in \mathbb{R}^+$ is a prescribed feasibility index, and $\sigma^{(j)} \in \mathbb{R}^+$ is an upper limit on the standard deviation of the $j$\textsuperscript{th} constraint. The mean input variables $\bar{\vec{s}}$ are bounded between lower and upper limits $\bar{\vec{s}}^{(l)}$ and $\bar{\vec{s}}^{(u)}$, respectively. The objective of RDO is to optimise the mean design variables $\bar{\vec{s}}$.

Input variables $\vec{s} =  \left (\vec{s}_d^\trans \,\, \vec{s}_f^\trans \right )^\trans$ of a computational model may be composed of immutable variables~\mbox{$\vec{s}_f \in \mathbb{R}^{d_f}$} that cannot be varied such as the PDE source terms, and design variables $\vec{s}_d \in \mathbb{R}^{d_d}$ such as CAD geometry parameters. In that case,  the mean input variables $\bar{\vec{s}}$ are bounded between lower and upper limits~$\bar{\vec{s}}^{(l)} = \left ((\bar{\vec{s}}_d^{(l)})^\trans\,\, \bar{\vec{s}}_f^\trans \right )^\trans$ and $\bar{\vec{s}}^{(u)}= \left ((\bar{\vec{s}}_d^{(u)})^\trans\,\, \bar{\vec{s}}_f^\trans \right )^\trans$, respectively, and the objective of RDO becomes to optimise the mean design variables $\bar{\vec{s}}_d$.

\subsection{Review of Gaussian processes}\label{subsection:gaussian_process_regression}

\begin{figure}[t!]
\centering
\includegraphics[width = 80mm]{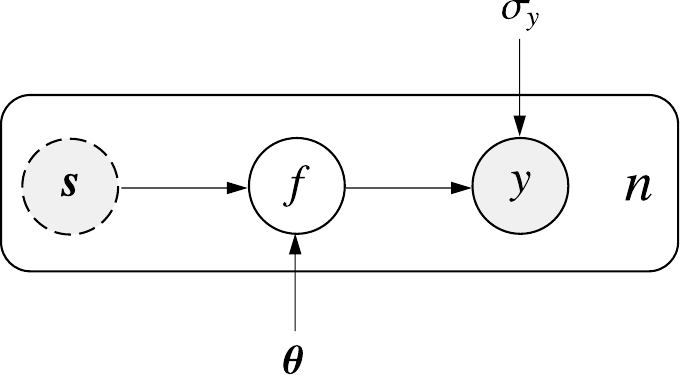}
\caption{Graphical model of a standard GP where $n$ is the size of the training data set $\mathcal{D}$, $\vec{s}$ is the vector of (deterministic) input variables, $f$ is the (random) target output variable and $y$ is the noisy observation. The model hyperparameters consist of  the prior hyperparameters $\vec{\theta}$ and the noise standard deviation $\sigma_y$.}
\label{fig:graphical_model_gp}
\end{figure}

In RDO, the objective $J(\vec{s})$ and constraint $H^{(j)}(\vec{s})$ functions are emulated by GP surrogates denoted as $f(\vec{s})$. A single observation from either function denoted $y \in \mathbb{R}$, follows a statistical observation model, see Figure \ref{fig:graphical_model_gp},
\begin{equation}\label{eq:gp1}
\begin{aligned}
y = f(\vec{s}) + \epsilon_y , \quad \quad \quad \quad \epsilon_y \sim \mathcal{N}(0,\sigma_y^2),
\end{aligned}
\end{equation}
where $\sigma_y \in \mathbb{R}^+$ denotes the noise or error standard deviation, such as the FE discretisation error, and the GP $f:\mathbb{R}^{d_s}\rightarrow\mathbb{R}$ maps between the input variables $\vec{s}$ and observation $y$, and has the prior
\begin{equation}\label{eq:gp2}
\begin{aligned}
f(\vec{s}) \sim \mathcal{GP}\left(\bar{f}(\vec{s})\,,\,c(\vec{s},\vec{s}')\right).
\end{aligned}
\end{equation}
Without loss of generality, we choose a zero-mean prior $\bar{f}(\vec{s}) = 0$ and squared exponential covariance function $c:\mathbb{R}^{d_s}\times\mathbb{R}^{d_s}\rightarrow\mathbb{R}$ (other types of kernels may be used when the objective is less smooth),
\begin{equation}\label{eq:gp3}
\begin{aligned}
c(\vec{s},\vec{s}') = \sigma_f^2\exp\left(-\sum_{i=1}^{d_s}\frac{(s_i-s_i')^2}{2 \ell_i}\right),
\end{aligned}
\end{equation}
where $\sigma_f \in \mathbb{R}^+$ is a scaling parameter, $\ell_i \in \mathbb{R}^+$ is a length scale parameter,  and $s_i$ denotes the $i$\textsuperscript{th} component of input vector $\vec{s}$. The hyperparameters of the GP are summarised as $\vec{\theta} = \{\sigma_f,\ell_1,\ell_2,...,\ell_{d_s}\}$. 

Let $\mathcal{D} = \{(\vec{s}_i,y_i)\,\vert\,  i= 1, 2, \dotsc ,  n \}$ denote the training data set, which is collected into the input variable matrix $\vec{S} = \left(\vec{s}_1 \,\, \vec{s}_2\,\,...\,\,\vec{s}_n\right)^\trans \in \mathbb{R}^{n \times d_s}$ and observation vector $\vec{y} = \left(y_1\,\,y_2\,\,...\,\,y_n\right)^\trans \in \mathbb{R}^n$. The joint probability density of observations $\vec{y}$ and target output variables $\vec{f} = \left(f_1\,\,f_2\,\,...\,\,f_n\right)^\trans \in \mathbb{R}^{n}$ can be factorised as
\begin{equation}\label{eq:gp4}
\begin{aligned}
p_\Theta(\vec{y},\vec{f}) = p_{\sigma_y}(\vec{y} \vert \vec{f})p_\theta(\vec{f}),
\end{aligned}
\end{equation}
which is expanded using the likelihood of assumed independent and identically distributed (iid) observations 
\begin{equation}\label{eq:gp5}
\begin{aligned}
p_{\sigma_y}(\vec{y}\vert\vec{f}) = \prod_{i=1}^n p_{\sigma_y}\left(y_i\vert f_i\right) = \mathcal{N}\left(\vec{f},\sigma_y^2\vec{I}\right),
\end{aligned}
\end{equation}
where $y_i$ and $f_i$ denote the observation and target output variable for the $i$\textsuperscript{th} training data sample, and the GP prior probability density
\begin{equation}\label{eq:gp6}
\begin{aligned}
p_{\theta}(\vec{f}) = \mathcal{N}(\vec{0},\vec{C}_{SS}).
\end{aligned}
\end{equation}
The entries of the covariance $\vec{C}_{SS} \in \mathbb{R}^{n \times n}$ consist of the covariance function $c(\vec{s},\vec{s}')$ where $\vec{s}$ and $\vec{s}'$ are rows of~$\vec{S}$. The marginal likelihood
\begin{equation}\label{eq:gp7}
\begin{aligned}
p_{\Theta}(\vec{y}) = \int p_{\sigma_y}(\vec{y} \vert \vec{f}) p_{\theta}(\vec{f})  \D \vec{f} = \mathcal{N}\left(\vec{0}\,,\,\vec{C}_{SS}+\sigma_y^2\vec{I}\right),
\end{aligned}
\end{equation}
is derived by marginalising over the target output variables $\vec{f}$. In empirical Bayes' methods, the optimum hyperparameters $\vec{\Theta} = \vec{\theta} \cup \{\sigma_y\}$ are computed by maximising the log marginal likelihood 
\begin{equation}\label{eq:gp8}
\begin{aligned}
\vec{\Theta}^* = \argmax_\Theta \ln p_\Theta(\vec{y}),
\end{aligned}
\end{equation}
where
\begin{equation}\label{eq:gp9}
\begin{aligned}
\ln p_\Theta(\vec{y}) = -\frac{n}{2}\ln(2\pi) -\frac{1}{2}\ln \left\vert\vec{C}_{SS}+ \sigma_y^2 \pmb{I} \right\vert -\frac{1}{2}\vec{y}^{\trans}\left(\vec{C}_{SS}+ \sigma_y^2 \pmb{I}\right)^{-1}\vec{y},
\end{aligned}
\end{equation}
and $\vert \cdot \vert$ denotes the determinant. For any new test input variables $\vec{S}_* \in \mathbb{R}^{n_* \times d_s}$, the posterior predictive probability density of the test output variables $\vec{f}_* \in \mathbb{R}^{n_*}$ is given by
\begin{equation}\label{eq:gp10}
\begin{aligned}
p_\Theta(\vec{f}_*\vert\vec{y}) = \int p_\theta(\vec{f}_*\vert\vec{f}) p_\Theta(\vec{f} \vert \vec{y})  \D \vec{f},
\end{aligned}
\end{equation}
which assumes conditional independence between the test output variables $\vec{f}_*$ and observations $\vec{y}$ when  target output variables $\vec{f}$ are given. The posterior probability densities $p_\theta(\vec{f}_*\vert\vec{f})$ and $p_\Theta(\vec{f} \vert \vec{y})$ can be determined using Bayes rule
\begin{equation}\label{eq:gp11}
\begin{aligned}
p_\theta(\vec{f}_*\vert\vec{f}) = \frac{p_\theta(\vec{f}\vert\vec{f}_*) p_\theta(\vec{f}_*)}{p_\theta(\vec{f})}, \quad \quad \quad p_\Theta(\vec{f} \vert \vec{y}) = \frac{p_{\sigma_y}(\vec{y}\vert\vec{f})p_\theta(\vec{f})}{p_\Theta(\vec{y})}.
\end{aligned}
\end{equation}
Alternatively because all involved probability densities are Gaussian, the posterior probability density $p_\Theta(\vec{f}_*\vert\vec{y})$ is derived from the joint probability density
\begin{equation}\label{eq:gp12}
\begin{aligned}
p_\Theta(\vec{f}_*,\vec{y})
=\mathcal{N}\begin{pmatrix}\begin{pmatrix}
\vec{0}\\
\vec{0}
\end{pmatrix},
\begin{pmatrix}
\vec{C}_{S_{\!*}S_{\!*}}  & \vec{C}_{S_{\!*}S}\\
\vec{C}_{SS_{\!*}} &  \vec{C}_{SS} + \sigma_y^2 \vec{I}
\end{pmatrix}
\end{pmatrix}.
\end{aligned}
\end{equation}
The posterior probability density of test output variables $\vec{f}_*$ is given by
\begin{equation}\label{eq:gp13}
\begin{aligned}
p_\Theta(\vec{f}_*\vert\vec{y}) = \mathcal{N}\left(\vec{C}_{S_{\!*}S}\left(\vec{C}_{SS} + \sigma_y^2\vec{I}\right)^{-1}\vec{y} \,,\,\vec{C}_{S_{\!*}S_{\!*}} - \vec{C}_{S_{\!*}S}\left(\vec{C}_{SS} + \sigma_y^2\vec{I}\right)^{-1}\vec{C}_{SS_{\!*}}\right),
\end{aligned}
\end{equation}
which is derived by conditioning the joint probability density $p_\Theta(\vec{f}_*,\vec{y})$ on observations $\vec{y}$. The evaluation of the posterior probability density becomes computationally expensive for large training data sets due to the inversion of the covariance matrix $\vec{C}_{SS}$, which involves $\mathcal{O}(n^3)$ operations.

So far the input variables $\vec{S}$ have been assumed deterministic; however, if they are random and sampled from a prescribed probability density $p(\vec{S})$, the posterior probability density of the target output variables
\begin{equation}\label{eq:gp14}
\begin{aligned}
p_\Theta(\vec{f} \vert \vec{y}) = \frac{p_{\sigma_y}(\vec{y}\vert\vec{f}) \int p_\theta(\vec{f} \vert \vec{S}) p(\vec{S}) \D \vec{S}}{p_\Theta(\vec{y})},
\end{aligned}
\end{equation}
becomes analytically intractable because the covariance $\vec{C}_{SS}$ of the GP prior $p_\theta(\vec{f}\vert\vec{S})$ depends non-linearly, according to \eqref{eq:gp3}, on the input variables $\vec{S}$. 

%% file: statistical_model.tex
\section{Statistical surrogate model}\label{section:statistical_model}

In this section, we generalise standard GP regression to random input variables with intrinsic dimensionality reduction using a latent variable formulation. This is later extended to a sparse formulation of the latent variables, which reduces the computational expense of performing inference with input variable uncertainty. Finally, an algorithm for training the model and metrics for measuring accuracy are provided.

\subsection{Reduced dimension variational Gaussian process}\label{subsection:statistical_latent_variable_model}

\begin{figure}[b!]
\centering
\includegraphics[width = 85mm]{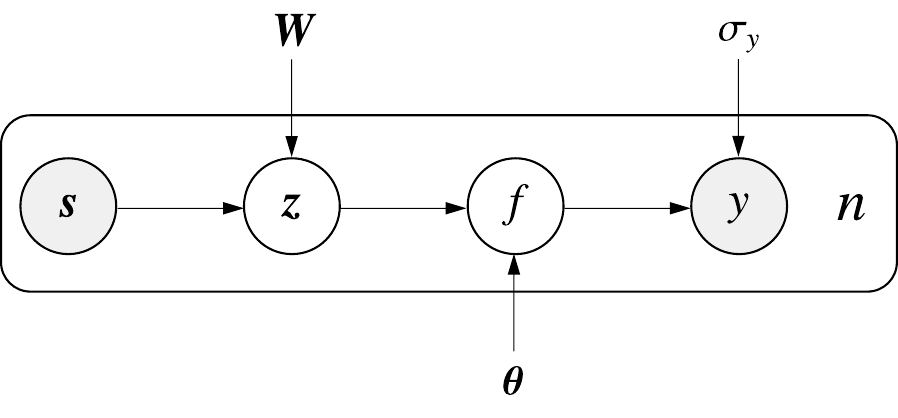}
\caption{Graphical model of the RDVGP surrogate where $n$ is the size of the training data set $\mathcal{D}$, $\vec{s}$ is the random input variable vector, $\vec{z}$ is the low-dimensional unobserved latent variable vector, $f$ is the target output variable and $y$ is the noisy observation. The model hyperparameters consist of the entries of the orthogonal projection matrix $\vec{W}$, the prior hyperparameters~$\vec{\theta}$ and the noise standard deviation~$\sigma_y$.}
\label{fig:graphical_model}
\end{figure}

We introduce a low-dimensional latent variable representation for each input variable $\vec{s}$, which is computed using the statistical observation model given by
\begin{subequations}\label{eq:plvm_1}
\begin{alignat}{2}
\vec{z} &= \vec{W}^\trans\vec{s}, \quad \quad\quad\quad&\vec{s} &\sim \mathcal{N}(\bar{\vec{s}},\vec{\Sigma}_s),\label{eq:plvm_1a}\\
y &= f(\vec{z}) + \epsilon_y, &\epsilon_y &\sim \mathcal{N}(0,\sigma_y^2).\label{eq:plvm_1b}
\end{alignat}
\end{subequations}
The low-dimensional latent variables $\vec{z} \in \mathbb{R}^{d_z}, \, (d_z < d_s)$ are mapped by a GP surrogate $f(\vec{z})$ to an observation $y$. The orthogonal projection matrix $\vec{W} \in \mathbb{R}^{d_s \times d_z}$ is composed of basis vectors of the low-dimensional subspace, and the mean $\bar{\vec{s}}$ and covariance $\vec{\Sigma}_s$ of the input variable probability density $p(\vec{s})$ are prescribed a priori. For a training data set $\mathcal{D}$ of size $n$, the joint probability density of the observations $\vec{y}$, target output variables $\vec{f}$, latent variables \mbox{$\vec{Z} = \left(\vec{z}_1\,\,\vec{z}_2\,\,...\,\,\vec{z}_n\right)^\trans \in \mathbb{R}^{n \times d_z}$}, and input variables $\vec{S}$, is factorised using the assumed conditional independence structure, see Figure \ref{fig:graphical_model}, such that
\begin{equation}\label{eq:plvm_2}
\begin{aligned}
p_\Theta(\vec{y},\vec{f},\vec{Z},\vec{S}) = p_{\sigma_y}(\vec{y} \vert \vec{f}) p_{\theta}(\vec{f} \vert \vec{Z}) p_{W}(\vec{Z} \vert \vec{S}) p(\vec{S}).
\end{aligned}
\end{equation}
The likelihood $p_{\sigma_y}(\vec{y} \vert \vec{f})$ is the same as in GP regression \eqref{eq:gp5}. The zero-mean GP prior probability density
\begin{equation}\label{eq:plvm_3}
\begin{aligned}
p_{\theta}(\vec{f} \vert \vec{Z}) = \mathcal{N}(\vec{0},\vec{C}_{ZZ}),
\end{aligned}
\end{equation}
has the covariance matrix $\vec{C}_{ZZ} \in \mathbb{R}^{n \times n}$ with the entries $c(\vec{z},\vec{z}')$ as defined in \eqref{eq:gp3}, where $\vec{z}$ and $\vec{z}'$ are rows of~$\vec{Z}$. The conditional probability density 
\begin{equation}\label{eq:plvm_4}
\begin{aligned}
p_W(\vec{Z}\vert\vec{S}) = \prod_{i=1}^n\delta\left(\vec{z}_i -\vec{W}^\trans\vec{s}_i\right),
\end{aligned}
\end{equation}
readily follows from \eqref{eq:plvm_1a}, where $\delta(\cdot)$ is the Dirac delta function. The input variable probability density
\begin{equation}\label{eq:plvm_5}
\begin{aligned}
p(\vec{S}) = \prod_{i=1}^n\mathcal{N}(\bar{\vec{s}}_i,\vec{\Sigma}_s),
\end{aligned}
\end{equation}
is prescribed a priori. We assume without loss of generality that the input variables are statistically independent from each other such that the covariance matrix~$\vec{\Sigma}_s$ is diagonal. The input variables $\vec{S}$ are marginalised out of the joint probability density, yielding
\begin{equation}\label{eq:plvm_6}
\begin{aligned}
p_\Theta(\vec{y},\vec{f},\vec{Z}) = p_{\sigma_y}(\vec{y} \vert \vec{f}) p_{\theta}(\vec{f} \vert \vec{Z}) p_{W}(\vec{Z}),
\end{aligned}
\end{equation}
where
\begin{equation}\label{eq:plvm_7}
\begin{aligned}
p_W(\vec{Z}) = \int p_W(\vec{Z}\vert\vec{S})p(\vec{S}) \D \vec{S} = \prod_{i=1}^n\mathcal{N}\left(\vec{W}^\trans\bar{\vec{s}}_i\,,\,\vec{W}^\trans\vec{\Sigma}_s\vec{W}\right).
\end{aligned}
\end{equation}
Although the input variables are statistically independent as implied by the measure of sampling, we note that the choice of projection matrix $\vec{W}$ induces interactions between input variables with respect to the outputs $\vec{y}$. The model hyperparameters \mbox{$\vec{\Theta} = \vec{\theta} \cup \{\vec{W},\sigma_y\}$} include in addition to the hyperparameters of the GP prior~$\vec{\theta}$, the noise or error standard deviation $\sigma_y$, and entries of the orthogonal projection matrix $\vec{W}$. Although the posterior probability density of all unobserved variables
\begin{equation}\label{eq:plvm_8}
\begin{aligned}
p_{\Theta}(\vec{f},\vec{Z} \vert \vec{y}) = \frac{p_\Theta(\vec{y},\vec{f},\vec{Z})}{p_\Theta(\vec{y})},
\end{aligned}
\end{equation}
follows from Bayes' rule, it is not available in closed-form. 

In VB, the posterior is approximated with a trial probability density $q_{\theta,\psi}(\vec{f},\vec{Z})$ where the KL divergence between the two probability densities is given by 
\begin{equation}\label{eq:plvm_9}
\begin{aligned}
D_{KL}\left(q_{\theta,\psi}(\vec{f},\vec{Z}) \,\vert \vert\, p_{\Theta}(\vec{f},\vec{Z}\vert\vec{y})\right) &= \int q_{\theta,\psi}(\vec{f},\vec{Z}) \ln \left(\frac{q_{\theta,\psi}(\vec{f},\vec{Z})}{p_{\Theta}(\vec{f},\vec{Z}\vert\vec{y})}\right) \D\vec{Z} \D\vec{f}\\
&= \int q_{\theta,\psi}(\vec{f},\vec{Z}) \ln \left(\frac{q_{\theta,\psi}(\vec{f},\vec{Z})p_\Theta(\vec{y})}{p_\Theta(\vec{y},\vec{f},\vec{Z})}\right) \D \vec{Z} \D \vec{f}\\
&= \int  q_{\theta,\psi}(\vec{f},\vec{Z}) \ln \left(\frac{q_{\theta,\psi}(\vec{f},\vec{Z})}{p_\Theta(\vec{y},\vec{f},\vec{Z})}\right) \D \vec{Z} \D \vec{f} + \ln p_\Theta(\vec{y})\\
&= -\mathcal{F}(\vec{y}) + \ln p_\Theta(\vec{y}).
\end{aligned}
\end{equation}
Since KL divergence is non-negative, i.e. $D_{KL}(\cdot \vert \vert \cdot) \geq 0$,
\begin{equation}\label{eq:plvm_10}
\begin{aligned}
\ln p_{\Theta}(\vec{y}) 
&\geq \int q_{\theta,\psi}(\vec{f},\vec{Z}) \ln \left(\frac{p_{\Theta}(\vec{y},\vec{f},\vec{Z})}{q_{\theta,\psi}(\vec{f},\vec{Z})}\right) \D \vec{Z} \D \vec{f}.
\end{aligned}
\end{equation}
The KL divergence is minimised by maximising the ELBO $\mathcal{F}(\vec{y})$. We assume a trial probability density 
\begin{equation}\label{eq:plvm_11}
\begin{aligned}
q_{\theta,\psi}(\vec{f},\vec{Z}) = p_\theta(\vec{f}\vert\vec{Z})q_\psi(\vec{Z}),
\end{aligned}
\end{equation}
with the Gaussian prior probability density $p_\theta(\vec{f}\vert\vec{Z})$ given by \eqref{eq:plvm_3}, and the trial probability density
\begin{equation}\label{eq:plvm_12}
\begin{aligned}
q_\psi(\vec{Z}) := \prod_{i=1}^{n}q_{\psi}(\vec{z}_i):=\prod_{i=1}^{n}\mathcal{N}\left(\tilde{\vec{\mu}}_{z,i}\,,\,\tilde{\vec{\Sigma}}_{z,i}\right).
\end{aligned}
\end{equation}
Variational parameters $\vec{\psi} = \left\{(\tilde{\vec{\mu}}_{z,i},\tilde{\vec{\Sigma}}_{z,i})\,\big\vert\, i= 1, 2, \dotsc ,  n \right\}$ consist of the components of the mean vector \mbox{$\tilde{\vec{\mu}}_{z,i} \in \mathbb{R}^{d_z}$} and the entries of the diagonal covariance matrix \mbox{$\tilde{\vec{\Sigma}}_{z,i} \in \mathbb{R}^{d_z \times d_z}$}. The ELBO is expanded by introducing the factorised joint probability density $p_\Theta(\vec{y},\vec{f},\vec{Z})$  and the factorised trial probability density $q_{\theta,\psi}(\vec{f},\vec{Z})$ as in~\eqref{eq:plvm_6} and~\eqref{eq:plvm_11} such that
\begin{equation}\label{eq:plvm_13}
\begin{aligned}
\mathcal{F}(\vec{y})
&= \int p_\theta(\vec{f}\vert\vec{Z})q_\psi(\vec{Z}) \ln \left(\frac{p_{\sigma_y}(\vec{y} \vert \vec{f}) p_{\theta}(\vec{f} \vert \vec{Z}) p_{W}(\vec{Z})}{p_\theta(\vec{f}\vert\vec{Z})q_\psi(\vec{Z})}\right) \D \vec{Z} \D \vec{f}\\
&= \int p_\theta(\vec{f}\vert\vec{Z})q_\psi(\vec{Z}) \left( \ln p_{\sigma_y}(\vec{y} \vert \vec{f}) + \ln \left(\frac{p_{W}(\vec{Z})}{q_\psi(\vec{Z})}\right) \right) \D \vec{Z} \D \vec{f}\\
&= \expect_{q_\psi(\vec{Z})}\left( \expect_{p_\theta(\vec{f}\vert\vec{Z})}\left( \ln p_{\sigma_y}(\vec{y} \vert \vec{f}) \right) \right) - D_{KL}\left(q_{\psi}(\vec{Z}) \, \vert \vert \, p_{W}(\vec{Z})\right).
\end{aligned}
\end{equation}
The KL divergence term in the ELBO is analytically tractable given that $q_\psi(\vec{Z})$ and $p_{W}(\vec{Z})$ belong to the exponential family of probability densities (Appendix \hyperlink{Appendix A}{A}). The expectations can be evaluated by MC sampling with $\vec{Z} \sim q_\psi(\vec{Z})$ and $\vec{f} \sim p_\theta(\vec{f}\vert\vec{Z})$. The ELBO $\mathcal{F}(\vec{y})$ can be extended to multiple (independent and identically distributed) observation vectors using a mean field approximation (Appendix \hyperlink{Appendix B}{B}). In downstream applications such as global optimisation, querying at $n_*$ test input variables can be computationally expensive for large sample sizes (when $n$ is large) because it requires $\mathcal{O}(n_*n^2)$ operations.
 
\subsection{Sparse formulation}\label{subsection:sparse_statistical_latent_variable_model}

\begin{figure}[t!]
	\centering
	\includegraphics[width = 85mm]{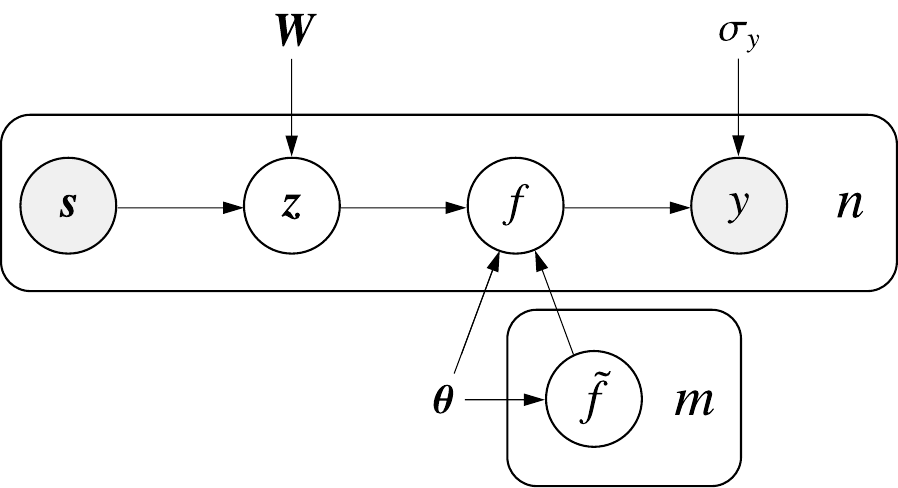}
	\caption{Graphical model of the sparse RDVGP surrogate where $n$ is the size of the training data set $\mathcal{D}$, $\vec{s}$ is the random input variable vector, $\vec{z}$ is the low-dimensional unobserved latent variable vector, $f$ is the target output variable, $\tilde{f}$ is the pseudo output variable with $m$ realisations and $y$ is the noisy observation. The model hyperparameters consist of the entries of the orthogonal projection matrix $\vec{W}$, the prior hyperparameters~$\vec{\theta}$ (which includes the pseudo latent variable $\tilde{\vec{z}}$) and the noise standard deviation~$\sigma_y$.}
	\label{fig:graphical_model_2}
\end{figure}

To reduce the computational expense of querying while retaining properties of the RDVGP surrogate, training data $\mathcal{D}$ is augmented with the $m$ (where $m \ll n$) pseudo output variables \mbox{$\tilde{\vec{f}} = (\tilde{f}_1\,\,\tilde{f}_2\,\,...\,\,\tilde{f}_m)^\trans \in \mathbb{R}^m$}. The pseudo output variables are trainable model parameters that when sampled at the pseudo latent variables \mbox{$\tilde{\vec{Z}} = \left(\tilde{\vec{z}}_1\,\,\tilde{\vec{z}}_2\,\,...\,\,\tilde{\vec{z}}_m \right)^\trans \in \mathbb{R}^{m \times d_z}$}, accurately explain the observations, see e.g. \cite{titsias2009variational}. The pseudo latent variables, also called inducing points, are considered as hyperparameters. The joint probability density $p_\Theta(\vec{y},\vec{f},\vec{Z})$ given by \eqref{eq:plvm_6} is augmented with the pseudo output variables $\tilde{\vec{f}}$ such that
\begin{equation}\label{eq:splvm_1}
\begin{aligned}
p_\Theta\left(\vec{y},\vec{f},\tilde{\vec{f}},\vec{Z}\right) = p_{\sigma_y}(\vec{y} \vert \vec{f}) p_{\theta}\left(\vec{f} \,\vert\, \tilde{\vec{f}}, \vec{Z}\right)p_{\theta}\left(\tilde{\vec{f}}\right) p_{W}(\vec{Z}),
\end{aligned}
\end{equation}
using the assumed conditional independence structure (Figure \ref{fig:graphical_model_2}). The new GP prior probability density
\begin{equation}\label{eq:splvm_2}
\begin{aligned}
p_{\theta}\left(\vec{f}\,\vert\, \tilde{\vec{f}} ,\vec{Z}\right) = \mathcal{N}\left(\vec{C}_{Z\tilde{Z}} \, \vec{C}_{\tilde{Z}\tilde{Z}}^{-1} \, \tilde{\vec{f}} \, , \, \vec{C}_{ZZ}-\vec{C}_{Z\tilde{Z}} \, \vec{C}_{\tilde{Z}\tilde{Z}}^{-1} \, \vec{C}_{\tilde{Z}Z}\right),
\end{aligned}
\end{equation}
is derived by conditioning the joint GP prior probability density $p_\theta(\vec{f},\tilde{\vec{f}}\,\vert \,\vec{Z}) = p_\theta(\vec{f}\,\vert \, \tilde{\vec{f}},\vec{Z}) p_\theta(\tilde{\vec{f}})$ on the pseudo output variables
$\tilde{\vec{f}}$, where
\begin{equation}\label{eq:splvm_3}
\begin{aligned}
p_\theta(\vec{f},\tilde{\vec{f}}\,\vert\,\vec{Z})
=\mathcal{N}\begin{pmatrix}\begin{pmatrix}
\vec{0}\\
\vec{0}
\end{pmatrix},
\begin{pmatrix}
\vec{C}_{ZZ}  & \vec{C}_{Z\tilde{Z}}\\
\vec{C}_{\tilde{Z}Z} &  \vec{C}_{\tilde{Z}\tilde{Z}}
\end{pmatrix}
\end{pmatrix}.
\end{aligned}
\end{equation}
This is similar to the derivation of standard GP regression, where $\vec{\theta} = \left\{\tilde{\vec{Z}},\sigma_f,\ell_1,\ell_2,...,\ell_{d_z}\right\}$ are hyperparameters of $p_\theta(\vec{f},\tilde{\vec{f}}\,\vert\,\vec{Z})$. The covariance $\vec{C}_{\tilde{Z}\tilde{Z}} \in \mathbb{R}^{m \times m}$ consists of the entries $c(\tilde{\vec{z}},\tilde{\vec{z}}')$, according to \eqref{eq:gp3},  where $\tilde{\vec{z}}$ and $\tilde{\vec{z}}'$ are rows of $\tilde{\vec{Z}}$. Similarly, the covariance $\vec{C}_{Z\tilde{Z}} \in \mathbb{R}^{n \times m}$ consists of entries $c(\vec{z},\tilde{\vec{z}})$ where $\vec{z}$ is a row of $\vec{Z}$. The posterior probability density of all unobserved variables
\begin{equation}\label{eq:splvm_4}
\begin{aligned}
p_{\Theta}\left(\vec{f},\tilde{\vec{f}},\vec{Z} \,\vert\, \vec{y}\right) = p_\theta\left(\vec{f}\,\vert\, \tilde{\vec{f}},\vec{Z},\vec{y}\right)p_\Theta\left(\tilde{\vec{f}}\,\vert\,\vec{Z},\vec{y}\right)p_{W}(\vec{Z},\vec{y}),
\end{aligned}
\end{equation}
is analytically intractable and is therefore approximated by a trial probability density
\begin{equation}\label{eq:splvm_5}
\begin{aligned}
q_{\theta,\Psi}\left(\vec{f}, \tilde{\vec{f}}, \vec{Z}\right) = p_\theta\left(\vec{f}\,\vert\, \tilde{\vec{f}},\vec{Z}\right) q_{\omega}\left(\tilde{\vec{f}}\right)q_{\psi}(\vec{Z}),
\end{aligned}
\end{equation}
where $\vec{\Psi} = \vec{\psi} \cup \vec{\omega}$ are variational parameters. We approximate the analytically intractable posterior probability density $p_\Theta(\tilde{\vec{f}}\, \vert \, \vec{y})$ with the Gaussian trial density 
\begin{equation}\label{eq:splvm_6}
\begin{aligned}
q_\omega\left(\tilde{\vec{f}}\right) := \mathcal{N}\left(\tilde{\vec{\mu}}_{\tilde{f}}\,,\,\tilde{\vec{\Sigma}}_{\tilde{f}}\right),
\end{aligned}
\end{equation}
where $\vec{\omega}=\{\tilde{\vec{\mu}}_{\tilde{f}},\tilde{\vec{\Sigma}}_{\tilde{f}}\}$ are the free variational parameters. The mean vector $\tilde{\vec{\mu}}_{\tilde{f}} \in \mathbb{R}^{m}$ and diagonal covariance matrix $\tilde{\vec{\Sigma}}_{\tilde{f}} \in \mathbb{R}^{m \times m}$ are trainable parameters.

For determining the hyperparameters and variational parameters, the KL divergence,  
\begin{equation}\label{eq:splvm_7}
\begin{aligned}
D_{KL}\left(q_{\theta,\Psi}\left(\vec{f},\tilde{\vec{f}},\vec{Z}\right) \,\vert \vert\, p_{\Theta}\left(\vec{f},\tilde{\vec{f}},\vec{Z}\,\vert\, \vec{y}\right)\right) 
&= \int  q_{\theta,\Psi}\left(\vec{f},\tilde{\vec{f}},\vec{Z}\right) \ln \left(\frac{q_{\theta,\Psi}(\vec{f},\tilde{\vec{f}},\vec{Z})}{p_\Theta(\vec{y},\vec{f},\tilde{\vec{f}},\vec{Z})}\right) \D \vec{Z} \D \tilde{\vec{f}} \D \vec{f} + \ln p_\Theta(\vec{y}),
\end{aligned}
\end{equation}
between the posterior probability density and Gaussian trial probability density is minimised. By factorising the KL divergence using the joint probability density $p_\Theta(\vec{y},\vec{f},\tilde{\vec{f}},\vec{Z})$ and trial probability density $q_{\theta,\Psi}(\vec{f},\tilde{\vec{f}},\vec{Z})$ as in \eqref{eq:splvm_1} and \eqref{eq:splvm_5}, respectively, we obtain the ELBO
\begin{equation}\label{eq:splvm_8}
\begin{aligned}
\mathcal{F}(\vec{y})
&= \expect_{q_\psi(Z)}\left( \hat{\mathcal{F}}(\vec{y},\vec{Z}) \right) - D_{KL}\left(q_{\omega}\left(\tilde{\vec{f}}\right) \,\vert \vert \,p_{\theta}\left(\tilde{\vec{f}}\right)\right) - D_{KL}\left(q_{\psi}(\vec{Z}) \,\vert \vert\, p_{W}(\vec{Z})\right),
\end{aligned}
\end{equation}
where $\hat{\mathcal{F}}(\vec{y})$ according to \cite{hensman2013gaussian}, is given by
\begin{equation}\label{eq:splvm_9}
\begin{aligned}
\hat{\mathcal{F}}(\vec{y},\vec{Z}) &= \expect_{q_\omega(\tilde{{f}})}\left( \expect_{p_\theta\left(f\,\vert\, \tilde{f},Z\right)} \left( \ln p_{\sigma_y}(\vec{y}\vert \vec{f}) \right) \right)\\
&= \ln \mathcal{N}\left(\vec{y} \,\vert\, \vec{C}_{Z\tilde{Z}} \, \vec{C}_{\tilde{Z}\tilde{Z}}^{-1} \, \tilde{\vec{\mu}}_{\tilde{f}} \,, \,\sigma_y^2\vec{I}\right) -\frac{1}{2\sigma_y^2}\left(\trace\left(\vec{C}_{ZZ}-\vec{C}_{Z\tilde{Z}} \, \vec{C}_{\tilde{Z}\tilde{Z}}^{-1} \, \vec{C}_{\tilde{Z}Z}\right)+\trace\left(\tilde{\vec{\Sigma}}_{\tilde{f}}\,\vec{C}_{\tilde{Z}\tilde{Z}}^{-1} \, \vec{C}_{\tilde{Z}Z} \, \vec{C}_{Z\tilde{Z}} \, \vec{C}_{\tilde{Z}\tilde{Z}}^{-1}\right)\right).
\end{aligned}
\end{equation}
MC sampling of $\vec{Z} \sim q_\psi(\vec{Z})$ is used to estimate the expectation in the ELBO $\mathcal{F}(\vec{y})$. The KL divergence terms are derived analytically given that all probability densities belong to the exponential family (Appendix \hyperlink{Appendix A}{A}).

The approximate posterior probability density of the test output variables $\vec{f}_*$ evaluated at test latent variables $\vec{Z}_* \in \mathbb{R}^{n_* \times d_z}$ given by
\begin{equation}\label{eq:splvm_10}
\begin{aligned}
q_{\theta,\omega}\left(\vec{f}_*\vert\vec{Z}_*\right) &= \int p_{\theta}\left(\vec{f}_*\,\vert\, \tilde{\vec{f}} ,\vec{Z}_*\right)q_\omega\left(\tilde{\vec{f}}\right) \D \tilde{\vec{f}}\\
&= \mathcal{N}\left( \vec{C}_{Z_*\tilde{Z}} \, \vec{C}_{\tilde{Z}\tilde{Z}}^{-1} \, \tilde{\vec{\mu}}_{\tilde{f}} \, , \, \vec{C}_{Z_*Z_*} - \vec{C}_{Z_*\tilde{Z}} \, \vec{C}_{\tilde{Z}\tilde{Z}}^{-1} \, \vec{C}_{\tilde{Z}Z_*} + \vec{C}_{Z_*\tilde{Z}} \, \vec{C}_{\tilde{Z}\tilde{Z}}^{-1} \, \tilde{\vec{\Sigma}}_{\tilde{f}} \, \vec{C}_{\tilde{Z}\tilde{Z}}^{-1} \, \vec{C}_{\tilde{Z}Z_*} \right),
\end{aligned}
\end{equation}
is derived by marginalising the conditional probability density \mbox{$p_{\theta}(\vec{f}_*\,\vert\, \tilde{\vec{f}} ,\vec{Z}_*)$} (of similar form to \eqref{eq:splvm_2}) over the pseudo output variables $\tilde{\vec{f}}$ sampled from the trial probability density $q_\omega(\tilde{\vec{f}})$. By augmenting the training data set $\mathcal{D}$ with pseudo output variables $\tilde{\vec{f}}$, the computational complexity of querying at $n_*$ test input variables is reduced from $\mathcal{O}(n_*n^2)$ to $\mathcal{O}(n_*m^2)$ (see also \cite{quinonero2005unifying,titsias2009variational,lawrence2007learning}), since the GP prior  $p_{\theta}(\vec{f}\vert\vec{Z}) = \int p_\theta(\vec{f} \,\vert\,\tilde{\vec{f}},\vec{Z}) p_\theta(\tilde{\vec{f}})\,  \D \tilde{\vec{f}}$ now depends on the covariance matrix $\vec{C}_{\tilde{Z}\tilde{Z}}$ over the $m$ pseudo latent variables $\tilde{\vec{Z}}$.

\subsection{Inference with input uncertainty}\label{subsection:inference}

It is computationally more efficient to optimise in RDO the RDVGP surrogate over the low-dimensional test latent variables $\vec{Z}_*$ instead of the high-dimensional test input variables $\vec{S}_*$. For any test latent variables~$\vec{Z}_*$ and corresponding test output variable $\vec{f}_*$, the approximate marginal posterior probability density given by
\begin{equation}\label{eq:i1}
\begin{aligned}
q_{\theta,\omega,W}\left(\vec{f}_*\right) = \int q_{\theta,\omega}\left(\vec{f}_*\vert \vec{Z}_*\right) p_W\left(\vec{Z}_*\right) \D \vec{Z}_*,
\end{aligned}
\end{equation}
is obtained from the derived approximate posterior probability density $q_{\theta,\omega}\left(\vec{f}_*\vert\vec{Z}_*\right)$ in \eqref{eq:splvm_10} and the marginal probability density
\begin{equation}\label{eq:i2}
\begin{aligned}
p_W(\vec{Z}_*) = \prod_{i_*=1}^{n_*}\mathcal{N}\left(\vec{W}^\trans\bar{\vec{s}}_{i_*}\,,\,\vec{W}^\trans\vec{\Sigma}_s\vec{W}\right),
\end{aligned}
\end{equation}
where the objective of RDO is to compute $\bar{\vec{z}}^* = \vec{W}^\trans\bar{\vec{s}}^{*}$, which is mapped to the input domain using the Moore-Penrose inverse $\vec{W}^\dagger$. For any test latent variable $\vec{z}_*$  (that is any row of $\vec{Z}_*$), the mean
\begin{equation}\label{eq:i3}
\begin{aligned}
\expect\left(f_*\right) = \expect_{p_W(z_*)}\left( \expect_{q_{\theta,\omega}(f_*\vert z_*) }\left(f_*\right)\right),
\end{aligned}
\end{equation}
and variance
\begin{equation}\label{eq:i4}
\begin{aligned}
\var\left(f_*\right) = \var_{p_W(z_*)}\left( \expect_{q_{\theta,\omega}(f_*\vert z_*) }\left(f_*\right)\right) + \expect_{p_W(z_*)}\left( \var_{p_{\theta,\omega}(f_*\vert z_*) }\left(f_*\right)\right),
\end{aligned}
\end{equation}
of the approximate marginal posterior probability density $q_{\theta,\omega,W}\left(f_*\right)$ given in \eqref{eq:i1}, can be approximated using the laws of total expectation and total variance, with MC sampling.

\subsection{Surrogate training} \label{subsection:model_training}

The RDVGP surrogate is trained by learning the set of hyperparameters $\vec{\Theta}= \vec{\theta} \cup \{\vec{W},\sigma_y\}$ and variational parameters $\vec{\Psi} = \vec{\psi} \cup \vec{\omega}$, that maximise the ELBO $\mathcal{F}(\vec{y})$ given by \eqref{eq:splvm_8}, such that
\begin{equation}\label{eq:mt1}
\begin{aligned}
\vec{\Theta}^*,\vec{\Psi}^* = \argmax_{\Theta,\Psi} \mathcal{F}(\vec{y}).
\end{aligned}
\end{equation}
We use the stochastic gradient method with the adaptive moment estimation algorithm (ADAM) \cite{kingma2014adam}, which evaluates the gradient of the expectation taken with respect to variational parameters of the latent variable trial probability density $q_\psi(\vec{Z})$  given by \eqref{eq:plvm_12}. Each ADAM iteration can be repeated $n_t$ times until convergence. We use the default ADAM step size and decay rates \cite{kingma2014adam}. As discussed in \cite{kingma2013auto}, to compute the gradient, the latent variable must use the reparameterisation given by
\begin{equation}\label{eq:mt2}
\begin{aligned}
\vec{z}_i = \tilde{\vec{\mu}}_{z,i} + \tilde{\vec{L}}_{z,i}\vec{\epsilon}_I,
\end{aligned}
\end{equation}
where $\vec{\epsilon}_I \sim \mathcal{N}(\vec{0},\vec{I})$ and $\tilde{\vec{L}}_{z,i}$ is the lower-triangular matrix computed using the Cholesky decomposition of $\tilde{\vec{\Sigma}}_{z,i}$ (this is known as the reparameterisation trick). The expectation in the ELBO $\mathcal{F}(\vec{y})$ is estimated using MC sampling with $n_l$ samples. We compute the ELBO gradient using a single MC sample estimate, which is sufficient for many practical applications \cite{kingma2019introduction}. 

The projection matrix $\vec{W}$ from the statistical observation model \eqref{eq:plvm_1a} diverges from the Stiefel manifold, i.e. is no-longer orthogonal, when using ADAM, therefore the Cayley ADAM algorithm \cite{li2020efficient} is used. We use the default step size and decay rates. An automatic relevance determination (ARD) procedure \cite{wipf2007new} is used to determine the dimension $d_z$ of the latent variables $\vec{z}$.

Furthermore, since the ELBO is non-convex, optimisation on the Stiefel manifold using gradient methods may converge to local minima \cite{horst2013handbook}, $n_r$ restarts with different initial hyperparameters are used (Algorithm~\ref{alg:rdvgp}). Exploration of highly redundant random input variables $\vec{s}$ can be prevented using a sparse orthogonal initialisation of the projection matrix $\vec{W}$; however, a random initialisation (using QR decomposition \cite{golub2013matrix}) is more appropriate for input variables $\vec{s}$ with low redundancy (we use a combination of both). The covariance matrices of the trial probability densities $\tilde{\vec{\Sigma}}_{\tilde{f}}$ and $\tilde{\vec{\Sigma}}_{z,i}$ are sampled from the space of positive definite diagonal matrices.

\begin{algorithm}[b!]
\caption{Pseudocode for training the RDVGP surrogate}\label{alg:rdvgp}
\begin{algorithmic}
\State {\small \textbf{Data}: $\mathcal{D} = \{(\vec{s}_i,y_i)\,\vert\, i= 1, 2, \dotsc ,  n\}$}\Comment{\small Sampled using LHS}
\State {\small \textbf{Input}: restarts $n_r$, ADAM iterations $n_t$, and MC samples $n_l$}
\For{\small $r \in \{1,2,...,n_r\}$}  
\Comment{Multiple restarts}
\State {\small Initialise model hyperparameters $\vec{\Theta}_0,\vec{\Psi}_0$}
\For{\small $t \in \{1,2,...,n_t\}$} \Comment{Gradient descent}
\For{\small $l \in \{1,2,...,n_l\}$} \Comment{MC sampling}
\State {\small $\vec{Z}_l \sim q_{\psi}(\vec{Z})$} \Comment{Reparameterisation \eqref{eq:mt2}}
\EndFor
\State {\small Compute $\mathcal{F}_t(\vec{y})$ and its gradient using $\{\vec{Z}_1,\vec{Z}_2,...,\vec{Z}_{n_l}\}$, $\vec{\Theta}_{t-1}$, and $\vec{\Psi}_{t-1}$} \Comment{ELBO \eqref{eq:splvm_8}}
\State{\small $\vec{\Theta}_t\backslash\{\vec{W}_t\},\vec{\Psi}_t \leftarrow \mathrm{ADAM}(\mathcal{F}_t(\vec{y}))$} \Comment{Stochastic gradient method}
\State{\small $\vec{W}_t \leftarrow \mathrm{CAYLEYADAM}(\mathcal{F}_t(\vec{y}))$ }  \Comment{Stiefel manifold optimisation}
\EndFor
\EndFor
\State {\small \textbf{Result}: optimum model hyperparameters $\vec{\Theta}^*,\vec{\Psi}^*$}
\end{algorithmic}
\end{algorithm}

The training data set $\mathcal{D}$ is determined using Latin hypercube sampling (LHS) \cite{mckay2000comparison}. Input variables \mbox{$\vec{s} = \left (\vec{s}_d^\trans \,\, \vec{s}_f^\trans \right )^\trans$} consist of immutable variables sampled from $\left\{\vec{s}_f \in \mathbb{R}^{d_f}\,\big\vert\, \bar{\vec{s}}_f-2\vec{\sigma}_{s_f} \leq \vec{s}_f \leq \bar{\vec{s}}_f+2\vec{\sigma}_{s_f}    \right\}$, $\bar{\vec{s}}_f$ and $\vec{\sigma}_{s_f}$ are vectors of the mean and standard deviations corresponding to each dimension. Similarly, the design variables are sampled from  $ \left\{\vec{s}_d \in \mathbb{R}^{d_d} \, \big\vert \,\bar{\vec{s}}_d^{(l)}-2\vec{\sigma}_{s_d} \leq \vec{s}_d \leq \bar{\vec{s}}_d^{(u)}+2\vec{\sigma}_{s_d} \right\}$, where $\bar{\vec{s}}_d^{(l)}$, $\bar{\vec{s}}_d^{(u)}$ and $\vec{\sigma}_{s_d}$ are vectors of lower limits, upper limits, and standard deviations corresponding to each dimension. 

\begin{figure*}[t!]
\centering
\subfloat[Slice sampling of $\expect_{p_\Theta(f_* \vert \vec{y},\vec{s}_*)} (f_*)$]{\includegraphics[width=81mm]{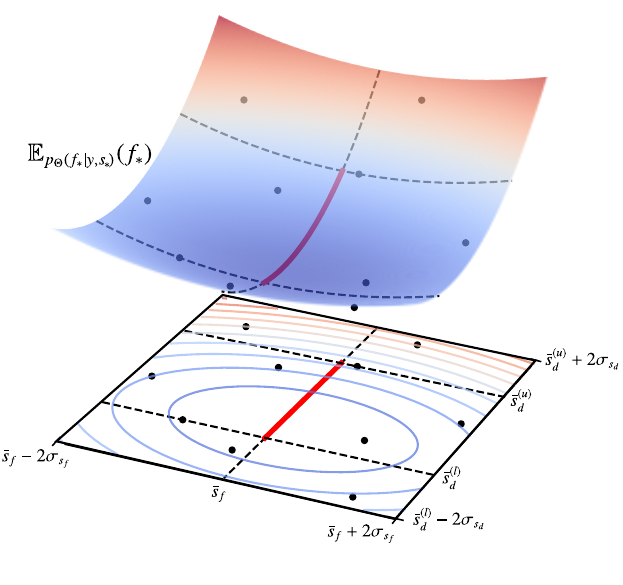}}
\centering
\subfloat[Marginal posterior density $p_\Theta(f_* \vert \vec{y})$ along slice $\bar{s}_*$]{\includegraphics[width=81mm]{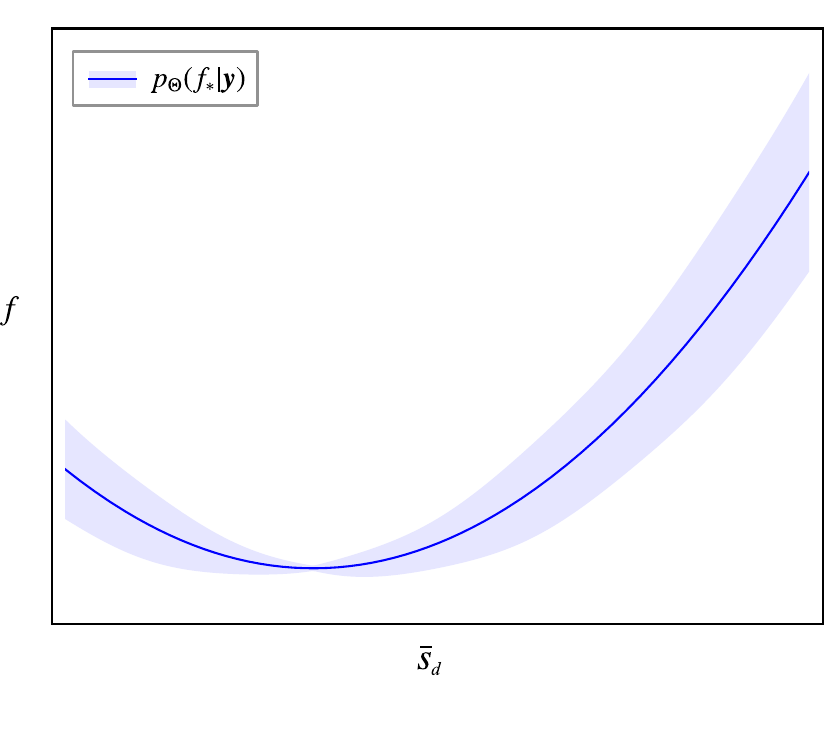}}
\caption{Schematic of slice sampling showing (a) the posterior probability density expected value $\expect_{p_\Theta(f_* \vert \vec{y},\vec{s}_*)} (f_*)$ along slice $\vec{s}_* = (s_{d*}\,\,\bar{s}_f)^\trans$, and (b) the marginal posterior probability density $p_\Theta(f_* \vert \vec{y})$ over mean test input variable $\bar{s}_* = (\bar{s}_{d*}\,\,\bar{s}_f)^\trans$.}
\label{fig:gp_slicing}
\end{figure*}

Slice sampling is used to infer the test output variables $\vec{f}_*$ corresponding to the design variables $\bar{\vec{s}}_{d*}$. The approximate posterior probability density $q_{\theta,\omega}(\vec{f}_*\vert\vec{z}_*)$ given by \eqref{eq:splvm_10}, is evaluated at the projected mean input variables $\bar{\vec{z}}_* = \vec{W}^\trans\bar{\vec{s}}_*$ where $\bar{\vec{s}}_* = \left (\bar{\vec{s}}_{d*}^\trans\,\,\bar{\vec{s}}_{f}^\trans \right)^\trans$ and $ \left\{\bar{\vec{s}}_{d*} \in \mathbb{R}^{d_d} \, \big\vert \,\bar{\vec{s}}_d^{(l)} \leq \bar{\vec{s}}_{d*} \leq \bar{\vec{s}}_d^{(u)}\right\}$, see Figure \ref{fig:gp_slicing}.

\subsection{Surrogate comparison}\label{subsection:model_comparison}

We verify the predictive accuracy of the RDVGP surrogate using two performance metrics; the coefficient of determination (COD) \cite{zhang2017coefficient}, and the maximum mean discrepancy (MMD) \cite{gretton2012kernel}. For any validation mean input variables $\bar{\vec{S}}_v$, corresponding true observations $\vec{y}_v$ of the objective or constraint function are obtained by sampling the input variable density $p(\vec{S}_v)$. Summary statistics of the true observations (such as mean and variance) can be compared to the inferred target output variables $\vec{f}_v$ . The COD values for the mean and variance
\begin{subequations}\label{eq:mc1}
\begin{alignat}{2}
R_\mu^2 &= 1 - \frac{\sum_{v=1}^{n_v}\big\vert \expect(\vec{y}_v)- \expect(\vec{f}_v)\big\vert^2}{\sum_{v=1}^{n_v}\big\vert \expect(\vec{y}_v)- \mu_{y_v}\big\vert^2},\label{eq:mc1a}\\
R_{\sigma}^2 &= 1 - \frac{\sum_{v=1}^{n_v}\big\vert \var(\vec{y}_v)- \var(\vec{f}_v)\big\vert^2}{\sum_{v=1}^{n_v}\big\vert \var(\vec{y}_v)- \sigma_{y_v}^2\big\vert^2}\label{eq:mc1b},
\end{alignat}
\end{subequations}
of the marginal posterior probability density $q_{\theta,\omega,W}(\vec{f}_v)$ in \eqref{eq:i1} are computed, where $\mu_{y_v}=\frac{1}{N_v}\sum_{v=1}^{n_v}\expect(\vec{y}_v)$ and $\sigma_{y_v}^2=\frac{1}{N_v}\sum_{v=1}^{n_v}\var(\vec{y}_v)$.

While the COD provides a measure of the quality of fit over the entire model, the distance between probability densities at points of interest such as local minima is important for determining the reliability of the predicted robust optimum design variables. The MMD $D_{MMD} \in \mathbb{R}^+$ provides this measure in terms of a mean embedding in reproducing kernel Hilbert space with associated kernel $k:\mathbb{R}^{d_y} \times \mathbb{R}^{d_y} \rightarrow \mathbb{R}$. For two probability densities $p(\vec{w})$ and $q(\vec{v})$, the MMD is
\begin{equation}\label{eq:mc2}
\begin{aligned}
D_{MMD}=\expect_{p(\vec{w})}\left(k(\vec{w},\vec{w}')\right) - 2\expect_{p(\vec{w}),q(\vec{v})}\left(k(\vec{w},\vec{v})\right) + \expect_{q(\vec{v})}\left(k(\vec{v},\vec{v}')\right).
\end{aligned}
\end{equation}
In all examples in this paper, we use a Gaussian kernel with unit amplitude and length scale to compute the MMD.

%% file: examples.tex
\section{Examples}\label{section:examples}

We demonstrate the accuracy of the proposed RDVGP surrogate using four examples of increasing complexity. In all examples, we use the sparse RDVGP surrogate, but refer to it as the RDVGP surrogate for brevity. Illustrative examples are used to compare the RDVGP surrogate to standard GP regression and to examine its versatility to different levels of input variance. We also demonstrate the RDVGP surrogate on RDO examples. The robust minimum as predicted by the RDVGP surrogate, is determined with genetic algorithms (NSGA-II), which use the emulated solutions inferred by the surrogate. To verify its accuracy, we directly MC sample the input variable and solve the corresponding FE model to estimate the true posterior probability density over the domain of the design variables. This is also optimised using genetic algorithms to yield an estimate of the true solution, for comparison.  

\subsection{One-dimensional illustrative example}\label{subsection:illustrative_example_1}

\begin{figure*}[t!]
	\centering
	\subfloat[Standard GP surrogate $p(\vec{J}_* \vert \vec{y}, \vec{s}_*)$]{\includegraphics[width=81mm]{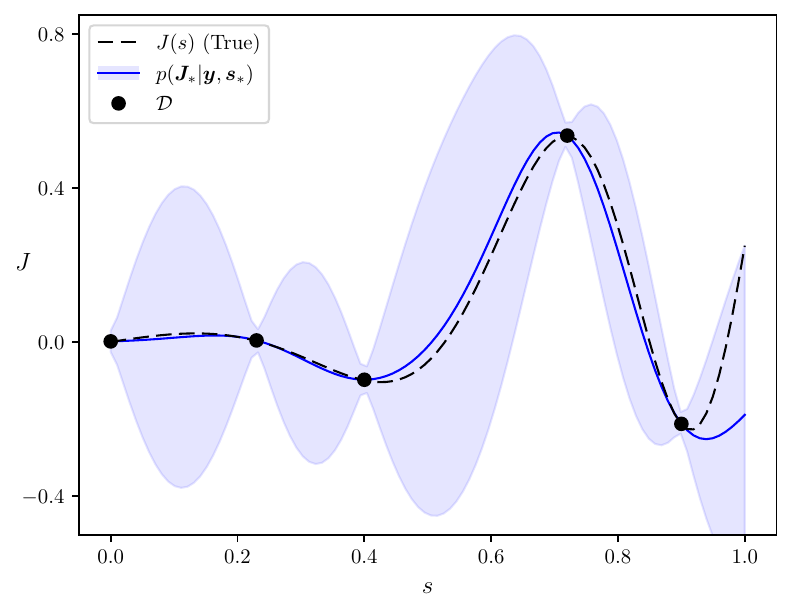}\label{fig:gp_f}}
	\centering
	\subfloat[Standard GP marginal posterior density $p(\vec{J}_* \vert \vec{y})$]{\includegraphics[width=81mm]{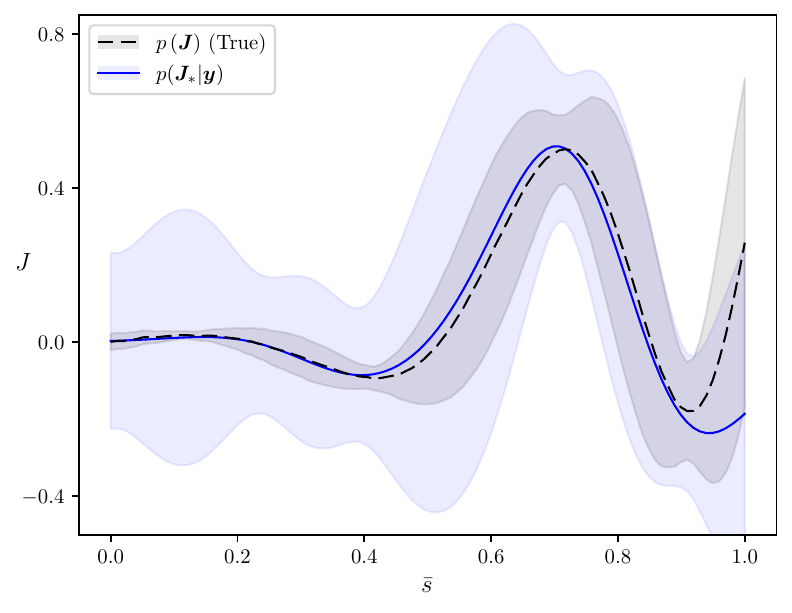}\label{fig:gp_y}}\\
	\centering
	\subfloat[RDVGP surrogate $q(\vec{J}_* \vert \vec{s}_*)$]{\includegraphics[width=81mm]{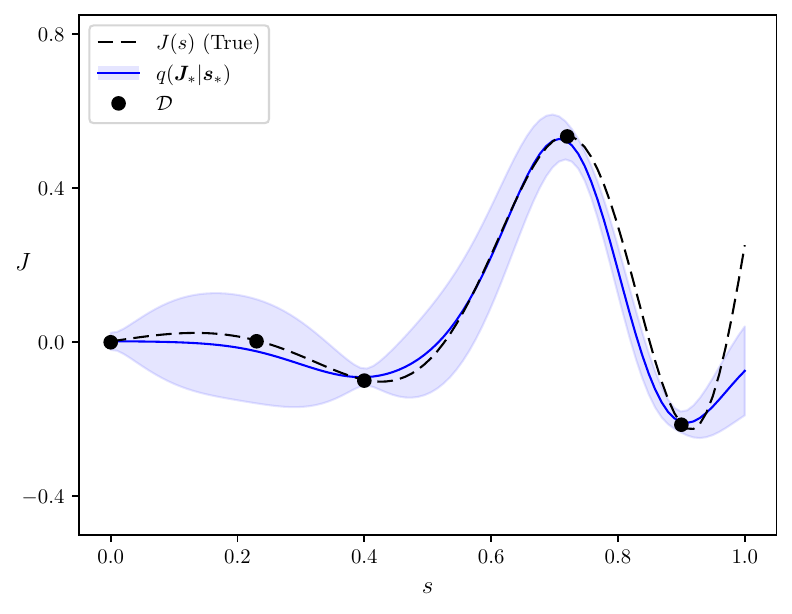}\label{fig:vi_f}}
	\centering
	\subfloat[RDVGP marginal posterior density $q(\vec{J}_*)$]{\includegraphics[width=81mm]{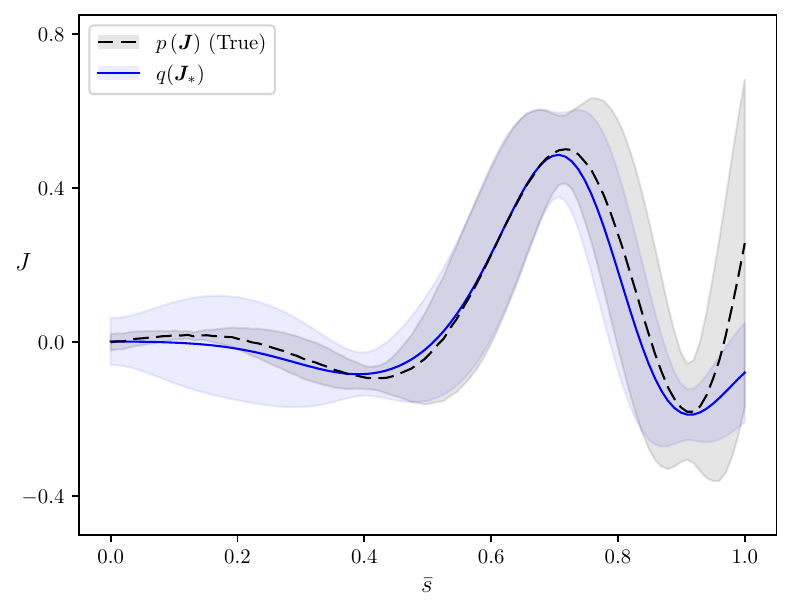}\label{fig:vi_y}}
	\caption{One-dimensional illustrative example. Comparing (a) the standard GP surrogate $p(\vec{J}_* \vert \vec{y}, \vec{s}_*)$ given by \eqref{eq:gp13} and (b) its marginal posterior probability density $p(\vec{J}_* \vert \vec{y})  =  \expect_{p(s_*)} \left( p(\vec{J}_* \vert \vec{y}, \vec{s}_*) \right)$, with (c) the RDVGP surrogate $q(\vec{J}_* \vert \vec{s}_*)$ given by \eqref{eq:splvm_10}, and (d) RDVGP marginal posterior probability density \mbox{$q(\vec{J}_*) = \expect_{p(s_*)} \left( q(\vec{J}_* \vert \vec{s}_*) \right)$} from \eqref{eq:i1}. The training data set $\mathcal{D}$ consists of $n=5$ random input variable $s$ and observation $\vec{y}$ pairs sampled from the true objective function $J(s)$, and $p(\vec{J})$ is the true marginal posterior probability density.}
	\label{fig:rdvgp_comparison}
\end{figure*}

The approximation of marginal posterior probability densities using the RDVGP surrogate is compared to the conventional approach of MC sampling from a standard GP surrogate as in \eqref{eq:gp13} \cite{zhou2020structural}. For both surrogates, the inferred variance depends on the random input variables $\vec{s}$, allowing variation over the input variable domain $\vec{s} \in D_s$. Consider the non-linear multi-modal objective function
\begin{equation}\label{eq:ea1}
\begin{aligned}
J(s) = -0.5s\sin{(3\pi s^2)} + 0.25s,
\end{aligned}
\end{equation}
where $s \sim \mathcal{N}(\bar{s},\sigma^2)$, $\bar{s} \in \{\bar{s} \in \mathbb{R} \, \vert \, 0 \leq \bar{s} \leq 1\}$, and the standard deviation  \mbox{$\sigma(\bar{s}) = 0.05 - 0.025\bar{s}$} is spatially varying. The training data set $\mathcal{D}$ is initialised using LHS with $n=5$ random samples (as shown in Figures~\ref{fig:gp_f} and \ref{fig:vi_f}) and the same number of inducing points (with $m=5$). This example considers only a single input variable, hence no dimensionality reduction is required. 
The standard GP and RDVGP surrogates are trained by maximising the log marginal likelihood $\ln p_\Theta(\vec{y})$ and ELBO $\mathcal{F}(\vec{y})$ given by \eqref{eq:gp9}  and \eqref{eq:splvm_8} respectively.

The standard GP overfits the training data (Figure \ref{fig:gp_f}), resulting in a posterior probability density $p(\vec{J}_*\vert \vec{y}, \vec{s}_*)$ with large epistemic uncertainty between sampled input variables. As mentioned in the introduction, epistemic uncertainty is caused by a lack of information and can in principle be reduced with additional training data. Conversely, aleatoric uncertainty is caused by the random input variable $\vec{s}$ itself \cite{roy2011comprehensive}. When using MC sampling to approximate the true marginal posterior probability density $p(\vec{J})$ using the standard GP posterior probability density $p(\vec{J}_*\vert \vec{y}, \vec{s}_*)$ given by \eqref{eq:gp13}, the aleatoric and epistemic uncertainties are confounded and the total uncertainty is significantly overestimated (Figure \ref{fig:gp_y}).

In contrast, the RDVGP surrogate does not overfit the training data (Figure \ref{fig:vi_f}); due to the inclusion of the prescribed input variable probability density $p(\vec{s})$ in the ELBO $\mathcal{F}(\vec{y})$, adding a data-informed regularisation. The expectation in the ELBO $\mathcal{F}(\vec{y})$ formulation promotes smoothness near the sampled training data, since each MC sample \mbox{$\vec{Z} \sim q_{\psi}(\vec{Z})$} (following from \eqref{eq:plvm_12}) is verified for a closeness of fit between the true observations $\vec{y}$ and the inferred target output variables $\vec{f}$ using a likelihood formulation $\hat{\mathcal{F}}(\vec{y},\vec{Z})$ given by \eqref{eq:splvm_9}, which promotes length scale hyperparameters that increase smoothness. This adds regularisation to the model and the use of pseudo target output variables $\tilde{\vec{f}}$ adds further flexibility when fitting the model (to prevent over-fitting). In comparison to the standard GP surrogate, this reduces epistemic uncertainty and more accurately estimates the total uncertainty, for improved selection of robust optimum design variables $\bar{\vec{s}}^*$ (Figure \ref{fig:vi_y}). In this example, the standard GP surrogate would require more training data than the RDVGP surrogate to reduce overfitting before accurately emulating the true marginal posterior probability density $p(\vec{J})$ (Appendix \hyperlink{Appendix C}{C}).

\subsection{Three-dimensional illustrative example}\label{subsection:illustrative_example_2}

This example investigates the accuracy and robustness of the RDVGP surrogate to different levels of input variable uncertainty and choices of parameters used to train the surrogate. Consider the minimisation of a non-linear multi-modal objective function
\begin{equation}\label{eq:e1}
\begin{aligned}
J(\vec{s}) = (s_2s_1-2)^2\sin{(12s_1-4)}+8s_1+s_3,
\end{aligned}
\end{equation}
of three dimensional input variables $\vec{s} = (s_1\,\,s_2\,\,s_3)^\trans$, subject to the constraint function
\begin{equation}\label{eq:e2}
\begin{aligned}
H(\vec{s}) = -\cos(2\pi s_1) - s_3 - 0.7,
\end{aligned}
\end{equation}
where $s_1 \sim \mathcal{N}(\bar{s}_1,\sigma_1^2)$ is a design variable, and $s_2 \sim \mathcal{N}(6.0,\sigma_2^2)$ and $s_3 \sim \mathcal{N}(0,0.1^2)$ are immutable variables. The mean of the design variable is sampled from the domain $\{\bar{s}_1 \in \mathbb{R}\,\vert\,0 \leq \bar{s}_1 \leq 1\}$. We perform RDO at three different choices for the constant variance, such that $\sigma_1\in\{0.025,0.05,0.075\}$ and $\sigma_2\in\{0.25,0.5,0.75\}$. According to the RDO problem stated in \eqref{eq:rdo1}, a weighting factor of $\alpha=0.25$ and feasibility index of $\beta = 2$ is chosen. No limit is imposed on the variance of the constraint function $H(\vec{s})$.

The training data set $\mathcal{D}$ is initialised using LHS with $n=100$ samples and choosing $m=25$ inducing points (reducing time complexity of inference by a factor of 16 from $\mathcal{O}(n_*n^2)$ to $\mathcal{O}(n_*m^2)$). 
The robust minimum variable $\bar{s}_1^*$ computed using the RDVGP surrogate is compared to the true minimum estimated using MC sampling $\langle\bar{s}_1^*\rangle$ with $10^{4}$ samples. A total of $n_v=30$ validation samples are used to compute COD values \eqref{eq:mc1} of the inferred objective and constraint marginal posterior probability densities $q(\vec{J}_*)$ and $q(\vec{H}_*)$ respectively. 

\begin{figure*}[!p]
		\centering
		{\includegraphics[width=150mm]{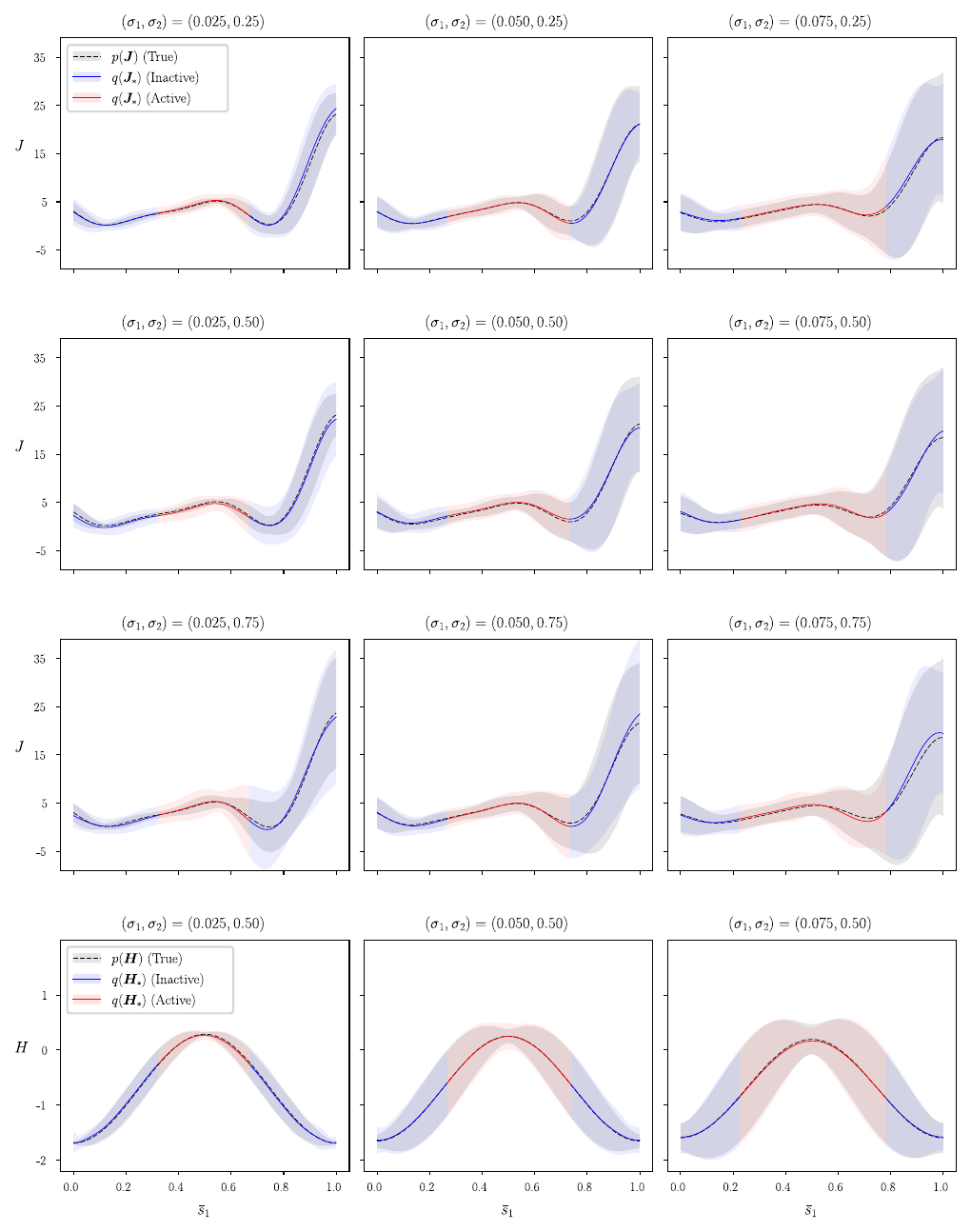}}\\
		\caption{Three-dimensional illustrative example. Comparison of the true $p(\cdot)$ and approximate (RDVGP) $q(\cdot)$ marginal posterior probability densities of the objective and constraint functions $J(\vec{s})$ and $H(\vec{s})$ over the domain of the design variable ${s}_1$. The design variable has density $\bar{s}_1 \sim \mathcal{N}(\bar{s}_1,\sigma_1^2)$ and the immutable variables have densities $\bar{s}_2 \sim \mathcal{N}(\bar{s}_2,\sigma_2^2)$ and $\bar{s}_3 \sim \mathcal{N}(\bar{s}_3,0.1^2)$. The constraint is active only in the red region.}
		\label{fig:ex1_posterior}
\end{figure*}

The immutable variable $s_3$ is substantially less influential than $s_1$ and $s_2$, giving a subspace dimension of $d_z=2$. An accurate robust minimum variable $\bar{s}_1^*$ is computed for all standard deviation combinations $\sigma_1$ and $\sigma_2$, when compared to the true solution (Table \ref{table:ex1_stats}). The largest computed percentage difference between the RDVGP $\bar{s}_1^*$ and true $\langle\bar{s}_1^*\rangle$ robust optimum design variables is 4\%. Although the objective function $J(\vec{s})$ has two local minima, one is not robust since the objective function varies significantly in close proximity to that local minima (Figure \ref{fig:ex1_posterior}). The approximate marginal posterior probability density $q(\vec{J}_*)$ computed using the RDVGP surrogate accurately emulates the true marginal posterior probability density $p(\vec{J})$, therefore the robust minimum $\bar{s}_1^*$ is well estimated. If the uncertainty in the random input variables becomes negligible, the global minimum would be $\bar{s}_1 = 0.749$, which does not coincide with the robust optimum design variable for any of the sampled statistics (Table \ref{table:ex1_stats}).

Large COD values $R_{\mu}^2$ are obtained from \eqref{eq:mc1a} for the mean of both the objective $J(\vec{s})$ and constraint $H(\vec{s})$ functions for all combinations of $\sigma_1$ and $\sigma_2$, indicating the mean is well predicted. However, the COD value $R_{\sigma}^2$ obtained from \eqref{eq:mc1b} for the variance of the objective and constraint functions  varies for different combinations of $\sigma_1$ and $\sigma_2$, and is most affected by the choice of $\sigma_1$. Furthermore, the MMD $D_{MMD}$, which compares the true $p(J(\langle\bar{s}_1^*\rangle))$ and approximate (RDVGP) $q(J(\langle\bar{s}_1^*\rangle))$ marginal posterior probability densities of the objective function at the true robust optimum design variable $\langle\bar{s}_1^*\rangle$ as given by \eqref{eq:mc2}, is affected by both $\sigma_1$ and $\sigma_2$. At smaller values for $\sigma_1$, the epistemic uncertainty makes up a greater proportion of the total uncertainty predicted by the RDVGP surrogate, increasing its discrepancy with the true solution. Since a slice of the full surrogate is taken at $s_2=6.0$, it becomes more difficult to approximate the aleatoric uncertainty at larger values of $\sigma_2$.

\begin{figure*}[t!]
\centering
\subfloat[MMD $D_{MMD}$ convergence]{\includegraphics[width=80mm]{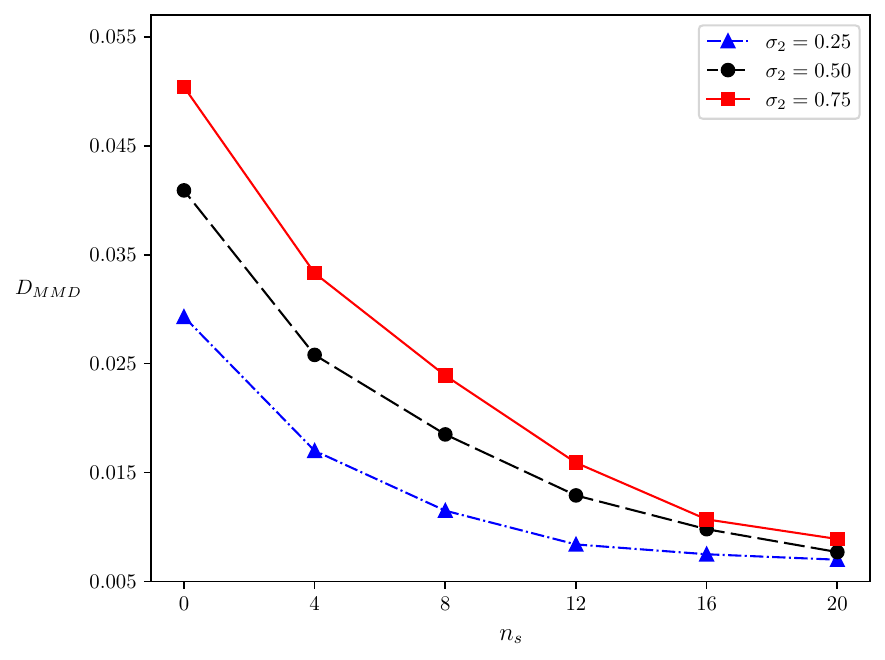}}
\centering
\subfloat[Optimum solution $\bar{s}_1^*$ convergence]{\includegraphics[width=80mm]{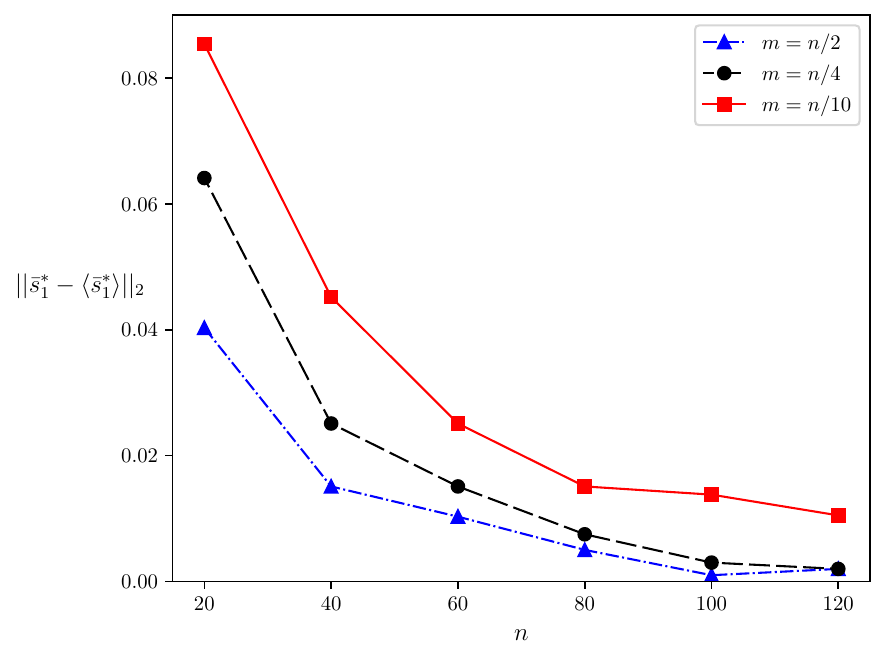}}
\caption{Three-dimensional illustrative example. In (a) the convergence plots show the effect of the number of supplementary training data points $n_s$ sampled within the vicinity of the true robust optimum design variable $\langle\bar{s}_1^*\rangle$, on the MMD $D_{MMD}$ computed between the true marginal posterior probability density $p(J(\langle\bar{s}_1^*\rangle))$ and the approximate (RDVGP) marginal posterior probability density $q(J(\langle\bar{s}_1^*\rangle))$ at $\sigma_1=0.025$ and varying values of $\sigma_2$. In (b) the convergence of the predicted solution $\bar{s}_1^*$ to the true robust optimum design variable $\langle\bar{s}_1^*\rangle$ for different training data sample sizes $n$ and choices of number of inducing points $m$, for \mbox{$(\sigma_1,\sigma_2) = (0.050,0.50)$} is shown.}
\label{fig:ex1_convergence_mmd}
\end{figure*}

As previously mentioned, the epistemic uncertainty can be reduced by increasing the size of training data set $\mathcal{D}$ (Figure \ref{fig:ex1_convergence_mmd}). By including supplementary data sampled using LHS centred on the true robust optimum design variable $\langle\bar{s}_1^*\rangle$ and box bounded within two standard deviations of each random input variable, the approximate (RDVGP) marginal posterior probability density $q(J(\langle\bar{s}_1^*\rangle))$ converges towards the true marginal posterior probability density $p(J(\langle\bar{s}_1^*\rangle))$ with an increasing number of supplementary samples $n_s$. This is demonstrated by the MMD $D_{MMD}$ decreasing for an increasing number of supplementary samples, a trend seen across all values of $\sigma_2$. Consequently, sequential or adaptive sampling strategies \cite{fuhg2021state} would be effective for reducing the epistemic uncertainty when using the RDVGP surrogate. 

\begin{table}[t!]
\small
\centering
\caption{Three-dimensional illustrative example. Robust minimum design variable as predicted by the RDVGP surrogate $\bar{s}_1^*$ and MC sampling $\langle\bar{s}_1^*\rangle$, with COD performance metrics computed using \eqref{eq:mc1} for the objective  \scalebox{0.9}{$\left(\left(R_\mu^J\right)^2,\left(R_\sigma^J\right)^2\right)$} and constraint \scalebox{0.9}{$\left(\left(R_\mu^H\right)^2,\left(R_\sigma^H\right)^2\right)$} functions, and MMD $D_{MMD}$ (computed using \eqref{eq:mc2}) between the approximate (RDVGP) $q(J(\langle\bar{s}_1^*\rangle))$ and true $p(J(\langle\bar{s}_1^*\rangle))$ marginal posterior probability densities evaluated at $\langle\bar{s}_1^*\rangle$.}\label{table:ex1_stats}
\begin{tabular}{c c c c c c c c c}
\toprule
$\sigma_1$ & $\sigma_2$ & $\left(R_\mu^J\right)^2$ & $\left(R_\sigma^J\right)^2$ & $\left(R_\mu^H\right)^2$ & $\left(R_\sigma^H\right)^2$ & $D_{MMD}$ & $\bar{s}_1^*$ & $\langle\bar{s}_1^*\rangle$\\ \midrule
\multirow{3}{*}{0.025} & 0.25 & 0.997 & 0.811 & 0.999 & 0.790 & 0.0293 & 0.128 & 0.123 \\ 
 & 0.50 & 0.993 & 0.828 & 0.999 & 0.789 & 0.0409 & 0.127 & 0.124 \\ 
 & 0.75 & 0.994 & 0.828 & 0.999 & 0.793 & 0.0504 & 0.127 & 0.124 \\ \midrule
\multirow{3}{*}{0.050} & 0.25 & 0.991 & 0.923 & 0.998 & 0.883 & 0.0175 & 0.137 & 0.138  \\ 
 & 0.50 & 0.985 & 0.924 & 0.999 & 0.888 & 0.0263 & 0.137 & 0.135\\
 & 0.75 & 0.988 & 0.929 & 0.998 & 0.887 & 0.0364 & 0.134 & 0.136 \\ \midrule
\multirow{3}{*}{0.075} & 0.25 & 0.992 & 0.959 & 0.995 & 0.949 & 0.0039 & 0.152 & 0.148\\ 
 & 0.50 & 0.974 & 0.965 & 0.995 & 0.959 & 0.0076 & 0.151 & 0.155\\ 
 & 0.75 & 0.969 & 0.944 & 0.994 & 0.932 & 0.0108 & 0.151 & 0.150\\ \bottomrule
\end{tabular}
\end{table}

\subsection{Bracket}\label{subsection:bracket}

\begin{figure*}[t!]
\centering
\subfloat[Domain with boundary conditions]{\includegraphics[width=80mm]{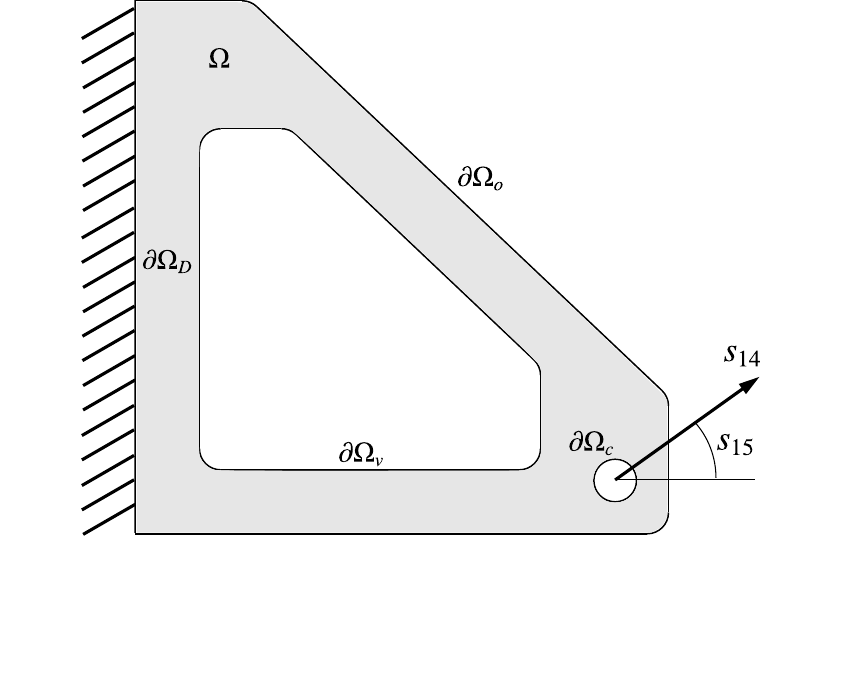}}
\centering
\subfloat[Bracket geometry]{\includegraphics[width=80mm]{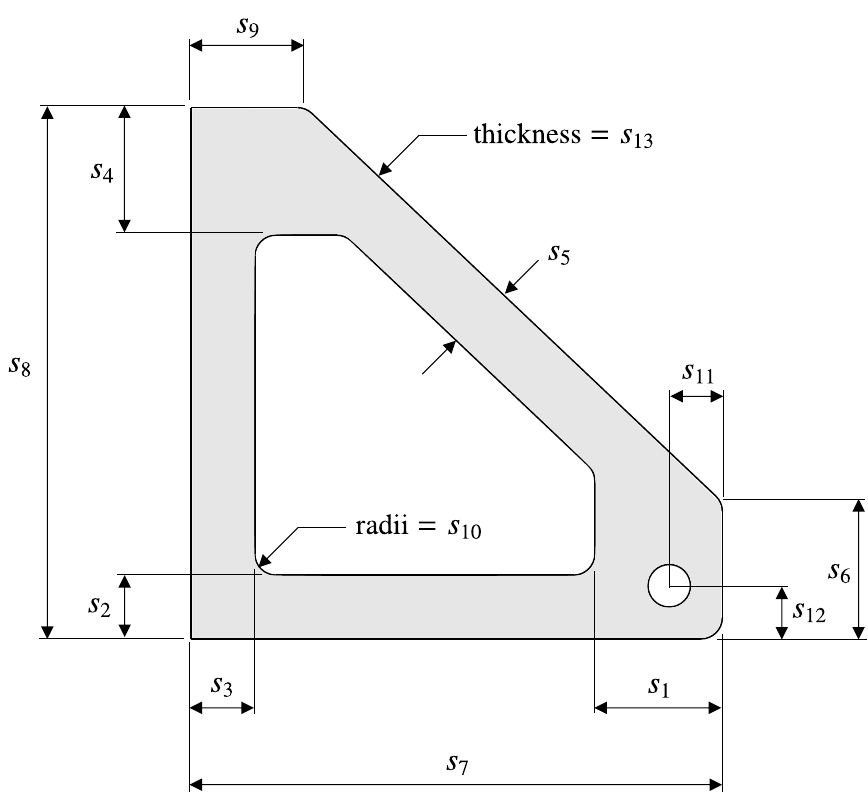}}
\caption{Bracket. Schematic showing (a) the domain of the bracket with a Dirichlet boundary condition applied on the left edge $\partial \Omega_D$, the Neumann boundary condition applied along the union of the outer boundary $\partial \Omega_o$, void boundary $\partial \Omega_v$, and circular hole boundary $\partial \Omega_c$ and (b) the bracket geometry with random design variables $\vec{s}_d = (s_1\,\,s_2\,\,...\,\,s_5)^\trans$ and random immutable variables $\vec{s}_f = (s_6\,\,s_7\,\,...\,\,s_{15})^\trans$.}
\label{fig:ex4_geometry}
\end{figure*}

Next, we demonstrate the use of the RDVGP surrogate on the RDO of the bracket shown in Figure~\ref{fig:ex4_geometry}. The deformation of the bracket is governed by linear elasticity equations. We denote $\vec{\sigma}_s\left(\vec{u}(\vec{x},\vec{s})\right) \in \mathbb{R}^{3 \times 3}$ as the stress tensor, $\vec{\varepsilon}\left(\vec{u}(\vec{x},\vec{s})\right) \in \mathbb{R}^{3 \times 3}$ as the strain tensor, $\vec{b}(\vec{x},\vec{s}) \in \mathbb{R}^3$ as the body force vector, $\vec{u}(\vec{x},\vec{s}) \in \mathbb{R}^3$ as the displacement vector, and $\vec{g}(\vec{x},\vec{s}) \in \mathbb{R}^3$ as the applied surface traction at the Neumann boundary $\partial \Omega_N$. The bracket has an outer boundary $\partial \Omega_o$, a polygonal void with boundary $\partial \Omega_v$, and a circular hole with boundary $\partial \Omega_c$. The Dirichlet boundary condition is applied along the left edge $\partial \Omega_D$, with the Neumann boundary condition applied along the remaining boundary $\partial \Omega_N$ where $\partial \Omega_N = \partial \Omega_o \cup \partial \Omega_v \cup \partial \Omega_c$.

The domain $\Omega$ is discretised over spatial coordinates $\vec{x} = (x_1\,\,x_2)^\trans$, and is parameterised by five design variables $\vec{s}_d = (s_1\,\,s_2\,\,...\,\,s_5)^\trans$ and ten immutable variables $\vec{s}_f = (s_6\,\,s_7\,\,...\,\,s_{15})^\trans$ controlling the shape of the domain $\Omega$, and the magnitude $s_{14}$ and angle $s_{15}$ of the resultant applied load. The body force $\vec{b}(\vec{x},\vec{s})$ is constant over the geometry domain, where $\vec{b}(\vec{x},\vec{s}) = (0\,\,-7.7\times 10^{-5}\,\, 0)^\trans$. The surface traction $\vec{g}(\vec{x},\vec{s})$ yields the resultant applied load (applied as uniformly distributed) along the circular hole boundary $\partial \Omega_c$ where \mbox{$\int_{\partial \Omega_c} \vec{g}(\vec{x},\vec{s}) \D  \vec{x} = (s_{14}\cos(s_{15})\,\,s_{14}\sin(s_{15})\,\,0)^\trans$}. All random input variables \mbox{$s_i \in \vec{s}$} are Gaussian, where \mbox{$s_i \sim \mathcal{N}(\bar{s}_i,\sigma_i^2)$} and $\bar{s}_i \in \mathbb{R}$ is a fixed constant for \mbox{$i\in \{6,7,...,15\}$}, and box-constrained on the interval \mbox{$ \bar{s}_i^{(l)} \leq \bar{s}_i \leq \bar{s}_i^{(u)}$} for $i\in\{1,2,...,5\}$ where $\bar{s}_i^{(l)}$ and $\bar{s}_i^{(u)}$ are the lower and upper limits respectively. 

The goal is to determine the robust optimum design variables $\bar{\vec{s}}_d^* = (\bar{s}_1^*\,\,\bar{s}_2^*\,\,...\,\,\bar{s}_5^*)^\trans$  that minimise structural compliance
\begin{equation}\label{eq:e10}
\begin{aligned}
J\left(\vec{s}\right) = \int_\Omega \vec{\sigma}_s\left(\vec{u}(\vec{x},\vec{s})\right) : \vec{\varepsilon}\left(\vec{u}(\vec{x},\vec{s})\right) \D \vec{x},
\end{aligned}
\end{equation}
subject to the volume constraint function
\begin{equation}\label{eq:e11}
\begin{aligned}
H^{(1)}(\vec{s}) = \int_\Omega \D \vec{x} - V_0,
\end{aligned}
\end{equation}
where $V_0=1.25\times10^5$ is the chosen limit on the volume. A constraint is also applied on the stress
\begin{equation}\label{eq:e12}
\begin{aligned}
H^{(2)}(\vec{s}) = \max_{x \in \Omega} \sigma_v(\vec{u}(\vec{x},\vec{s})) - \sigma_{v0},
\end{aligned}
\end{equation}
where $\sigma_v(\vec{u}(\vec{x},\vec{s}))$ is the von Mises stress, and $\sigma_{v0}=125$ is the upper limit on the stress. According to the RDO formulation \eqref{eq:rdo1}, a weighting factor of $\alpha=0.5$ and feasibility indices of $\beta^{(1)}=\beta^{(2)}=2$ are used. The standard deviation in the maximum von Mises stress $\sigma_v(\vec{u}(\vec{x},\vec{s}))$ must also be less than $35$.

\begin{figure*}[b!]
	\centering
	{\includegraphics[width=80mm]{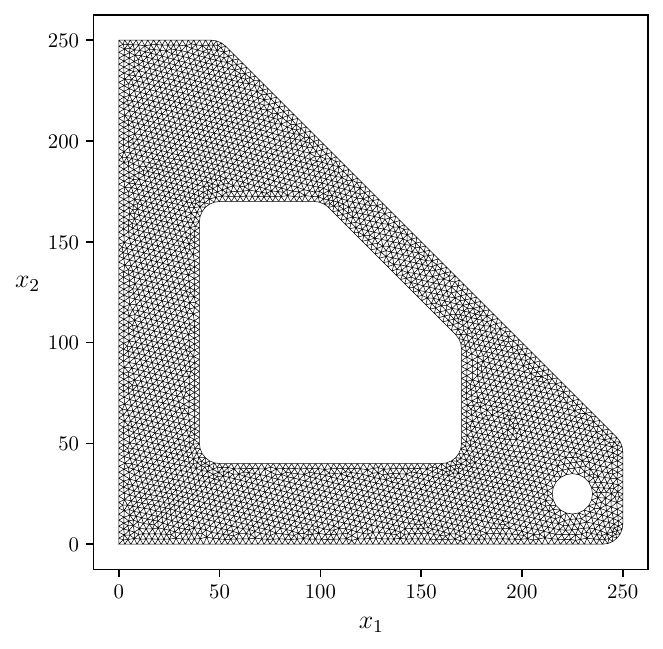}}
	\caption{Bracket. FE mesh of an example geometry containing 7668 linear triangle elements, discretised over spatial coordinates $\vec{x} = (x_1\,\,x_2)^\trans$ according to the plane stress assumption.}
	\label{fig:ex4_mesh}
\end{figure*}

The training data set $\mathcal{D}$ is initialised using LHS with $n=500$ samples and $m=250$ inducing points.
The objective $J(\vec{s})$ and stress constraint $H^{(2)}(\vec{s})$ are sampled from a deterministic FE model (Figure \ref{fig:ex4_mesh}). To avoid sampling stress singularities, the maximum von Mises stress $\max \sigma_v(\vec{u}(\vec{x},\vec{s}))$ is not sampled within a distance of 5 from the Dirichlet $\partial\Omega_D$ and Neumann $\partial\Omega_N$ boundary condition surfaces. The bracket is modelled as a thin plate, using the plane stress assumption. A Young's modulus of $E = 2.1\times10^{5}$  and Poisson's ratio of $\upsilon = 0.3$ are used. The mean and standard deviation of the immutable variables $\vec{s}_f$ are prescribed. Similarly, the lower and upper bounds on the mean, and standard deviation of the design variables $\vec{s}_d$ are also prescribed (Table \ref{table:ex4_bounds}). The normalisation constants $\bar{\mu}$ and $\bar{\sigma}$ in the weighted sum \eqref{eq:rdo1} are set equal to the mean and standard deviation of the objective function $J(\vec{s})$ observations. 

The robust optimum design variables $\bar{\vec{s}}_d^*$ computed using the RDVGP surrogate are compared to the true optimum $\langle\bar{\vec{s}}_d^*\rangle$ evaluated using MC sampling with $10^5$ samples. A total of $n_v=30$ validation samples are used for computing COD values given by \eqref{eq:mc1}. The immutable variables $\vec{s}_f$ with negligible variance when propagated through the objective $J(\vec{s})$ and constraint $H^{(j)}(\vec{s})$ functions, contribute insignificantly to variance in the observations. This is a reasonable representation of many complex engineering systems manufactured with high-precision equipment. As expected, the subspace is primarily composed of a linear combination of the design variables $\vec{s}_d$ and the magnitude $s_{14}$ and angle $s_{15}$ of the applied load, where the projection matrix $\vec{W}$ has dimension $d_z = 7$.

\begin{table}[b!]
	\small
	\centering
	\caption{Bracket. Mean $\bar{s}_i$ and standard deviation $\sigma_i$ of the random input variables $\vec{s} = (s_1\,\,s_2\,\,...\,\,s_{15})^\trans$, where box constraints $\bar{s}_i \in [\bar{s}_i^{(l)},\bar{s}_i^{(u)}]$ are given for the design variables $\vec{s}_d = (s_1\,\,s_2\,\,...\,\,s_5)^\trans$.}\label{table:ex4_bounds}
	\scalebox{0.9}{
		\begin{tabular}{cccccccccccccccc}
			\toprule
			& $s_1$ & $s_2$ & $s_3$ & $s_4$ & $s_5$ & $s_6$ & $s_7$ & $s_8$ & $s_9$ & $s_{10}$ & $s_{11}$ & $s_{12}$ & $s_{13}$ & $s_{14}$ & $s_{15}$ \\\midrule
			$\bar{s}_i$ & [60,100] & [10,50] & [10,50] & [60,100] & [0,30] & 50 & 250 & 250 & 50 & 10 & 25 & 25 & 5 & 5000 & 0 \\
			$\sigma_i$ & 0.1 & 0.1  & 0.1 & 0.1 & 0.1 & 0.1 & 0.1 & 0.1 & 0.1 & 0.1 & 0.1 & 0.1 & 0.1  & 1000 & $\pi\slash6$\\
			\bottomrule
		\end{tabular}
	}
\end{table}

\begin{figure*}[t!]
\centering
\subfloat[Stress mean $\expect\left(\sigma_v(\vec{x},\vec{s}^*)\right)$]{\includegraphics[width=70mm]{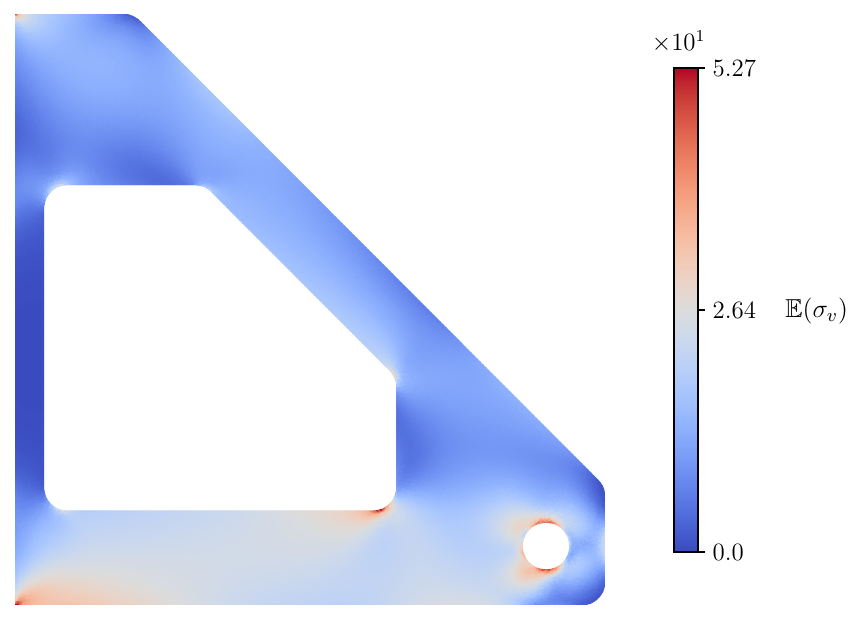}}
\centering
\subfloat[Stress std. dev. $\sqrt{\var\left(\sigma_v(\vec{x},\vec{s}^*)\right)}$]{\includegraphics[width=70mm]{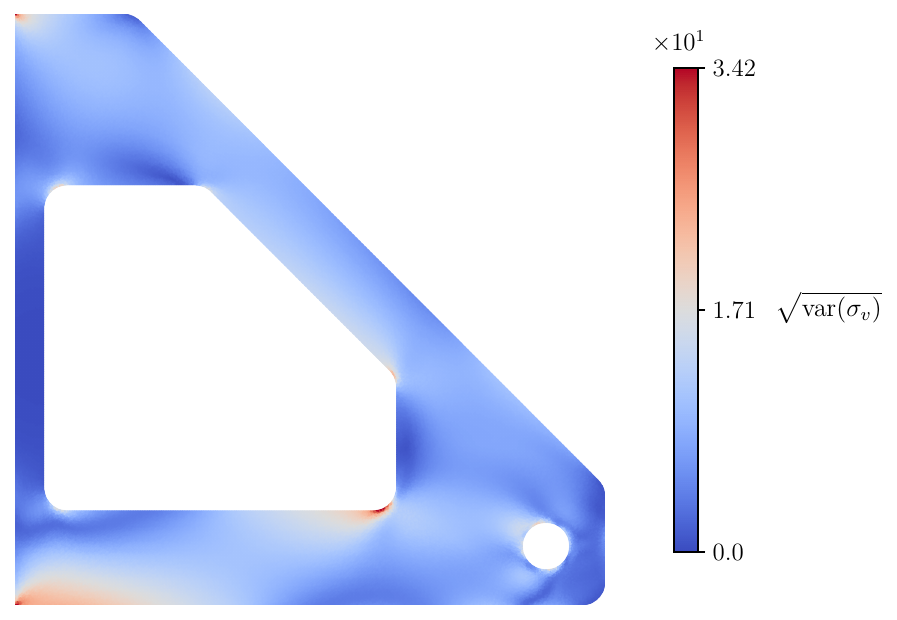}}\\
\subfloat[Displacement mean $\expect\left(\vert\vert\vec{u}(\vec{x},\vec{s}^*)\vert\vert_2\right)$]{\includegraphics[width=70mm]{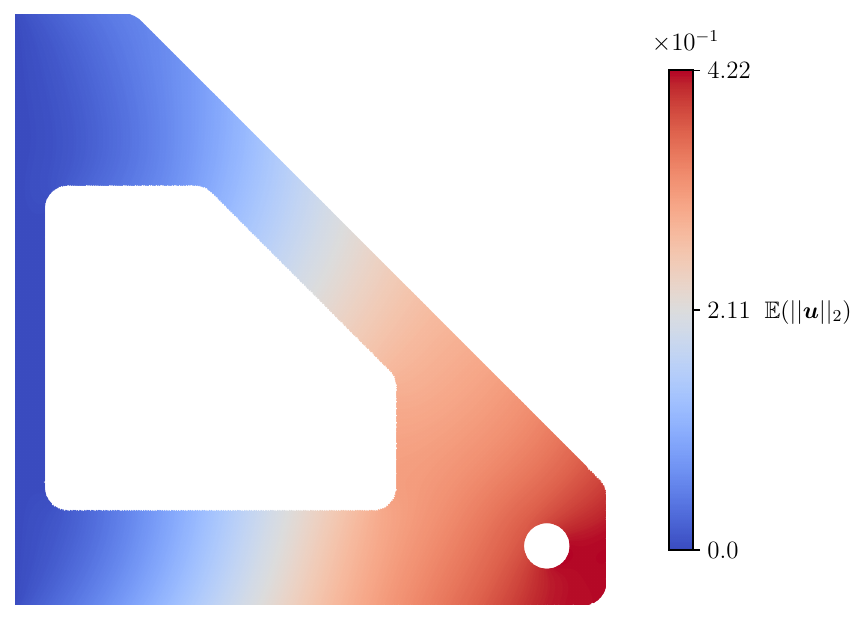}}
\centering
\subfloat[Displacement std. dev. $\sqrt{\var\left(\vert\vert\vec{u}(\vec{x},\vec{s}^*)\vert\vert_2\right)}$]{\includegraphics[width=70mm]{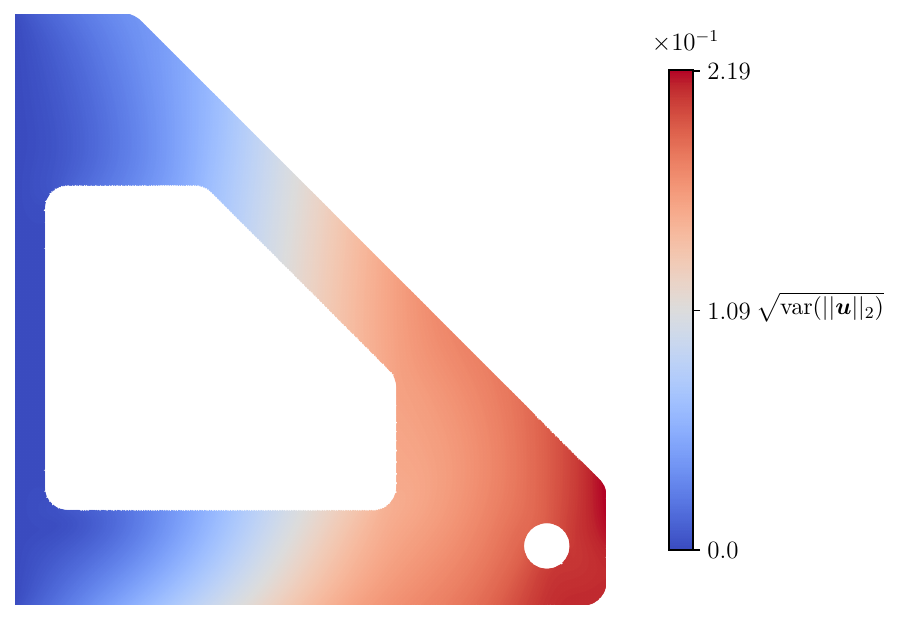}}
\caption{Bracket. Deformation showing for the optimum geometry $\vec{s}^* = ((\bar{\vec{s}}_d^*)^\trans\,\,\vec{s}_f^\trans)^\trans$ (a) the von Mises stress mean $\expect\left(\sigma_v(\vec{x},\vec{s}^*)\right)$, (b) the von Mises stress standard deviation $\sqrt{\var\left(\sigma_v(\vec{x},\vec{s}^*)\right)}$, (c) the displacement mean $\expect\left(\vert\vert\vec{u}(\vec{x},\vec{s}^*)\vert\vert_2\right)$, and (d) the displacement standard deviation $\sqrt{\var\left(\vert\vert\vec{u}(\vec{x},\vec{s}^*)\vert\vert_2\right)}$.}
\label{fig:ex4_res}
\end{figure*}

The true robust optimum design variables $\langle\bar{\vec{s}}_d^*\rangle=(88.6\,\,41.1\,\,10.5\,\,69.6\,\,29.2)^\trans$(as estimated with MC sampling) are well approximated by the robust optimum design variables \mbox{$\bar{\vec{s}}_d^*=(84.8\,\,40.2\,\,10.8\,\,75.2\,\,28.5)^\trans$} proposed by performing RDO \eqref{eq:rdo1} with the RDVGP surrogate. Due to the uncertainty in the magnitude $s_{14}$ and angle $s_{15}$ of the applied load (Figure \ref{fig:ex4_geometry}), the marginal posterior probability densities for the compliance $q(\vec{J}_*)$ and the stress constraint $q(\vec{H}_*^{(2)})$ yield a large aleatoric uncertainty prediction (Figure \ref{fig:ex4_res}). Consequently, permissible COD values are obtained for the mean $R_\mu^2$ (0.801,0.907) and variance $R_\sigma^2$ (0.757,0.716) of the objective $J(\vec{s})$ and stress constraint $H^{(2)}(\vec{s})$ functions respectively. Since the aleatoric uncertainty in the volume constraint $H^{(1)}(\vec{s})$ is negligible due to the small standard deviation in the random input variables $\vec{s}$ (Table \ref{table:ex4_bounds}), the epistemic uncertainty constitutes most of the total variance, and the RDVGP surrogate becomes a poor predictor of the aleatoric uncertainty in the volume constraint $H^{(1)}(\vec{s})$. This is demonstrated by COD values of 0.864 and 0.331 for the mean and variance respectively. The probability density near the robust optimum design variables is accurate, as demonstrated by an MMD $D_{MMD}$ (computed using \eqref{eq:mc2}) of 0.000983  between the true $p(J(\langle\bar{\vec{s}}_d^*\rangle))$ and approximate (RDVGP) $q(J(\langle\bar{\vec{s}}_d^*\rangle))$ marginal posterior probability densities. This example is a good demonstration of the RDVGP surrogate accuracy in uncertainty quantification with application to RDO.

\subsection{Cellular beam}\label{subsection:cellular_beam}
	
\begin{figure*}[b!]
	\centering
	\subfloat[Boundary conditions]{\includegraphics[width=130mm]{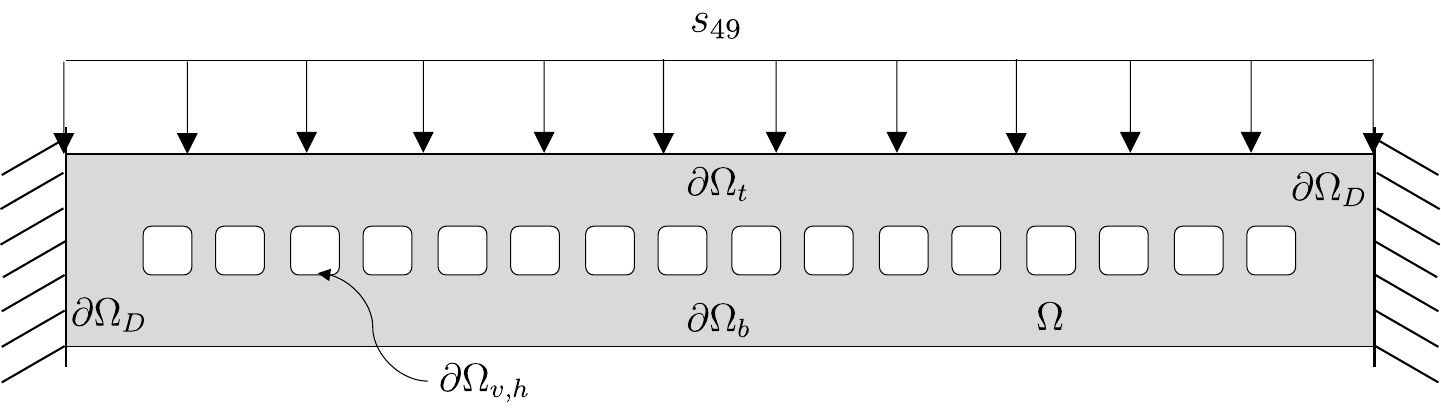}}\\
	\centering
	\subfloat[Design variables of the voids]{\includegraphics[width=50mm]{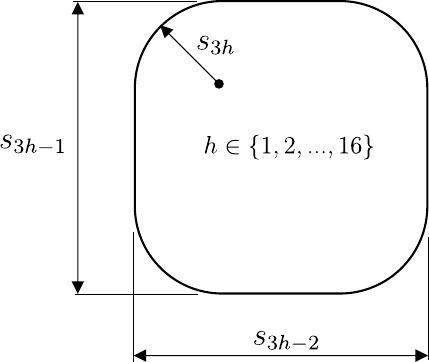}}
	\caption{Cellular beam. Schematic showing (a) the domain $\Omega$ of the beam with the Dirichlet boundary condition applied on the left and right edge $\partial \Omega_D$ and Neumann boundary conditions applied along the union of the top $\partial \Omega_t$, bottom $\partial \Omega_b$, and void $\partial \Omega_{v,h}$ (with $h \in \{1,2,...,16\}$), and (b) the void geometry with the design variables width $s_{3h-2}$, height $s_{3h-1}$ and fillet radius $s_{3h}$.}
	\label{fig:ex5_geo}
\end{figure*}

As an example with a large number of design variables and data points, we investigate the RDO of the linear elastic thin-rectangular cellular beam shown in Figure \ref{fig:ex5_geo}. The design of cellular beams is a frequently studied problem in structural optimisation~\cite{tsavdaridis2015application,martin2017analytical,rocha2023numerical}.  The domain of the beam $\Omega$ has a top $\partial \Omega_t$ and bottom $\partial \Omega_b$ boundary, and contains 16 uniformly spaced voids with boundary $\partial \Omega_{v,h}$, where \mbox{$h \in \{1,2,...,16\}$}. The Dirichlet boundary condition consisting of a fixed displacement constraint is applied along the left and right edges $\partial \Omega_D$, with the Neumann boundary condition consisting of a uniformly distributed load is applied along the remaining boundary $\partial \Omega_N$ where $\partial \Omega_N = \partial \Omega_t \cup \partial \Omega_b \cup \partial \Omega_{v,1} \cup ... \cup \partial \Omega_{v,16}$.

The domain $\Omega$ is discretised over spatial coordinates $\vec{x} = (x_1\,\,x_2)^\trans$, and is parameterised by 48 design variables \mbox{$\vec{s}_d = (s_1\,\,s_2\,\,...\,\,s_{48})^\trans$} and a single immutable variable $s_f = s_{49}$ which represents the magnitude of the uniformly distributed pressure applied along the top boundary $\partial \Omega_t$. There are 16 voids in total with each void having three design variables: a width $s_{3h-2}$, a height $s_{3h-1}$, and a fillet radius $s_{3h}$, which all control the shape of the geometry $\Omega$. The body force $\vec{b}(\vec{x},\vec{s}) = (0\,\,-7.7\times10^{-5}\,\,0)^\trans$ is constant over the domain. The surface traction $\vec{g}(\vec{x},\vec{s})$ is zero along the bottom and void boundaries  $\partial \Omega_b$ and $\partial \Omega_{v,h}$, respectively, and yields the resultant applied load  $\int_{\partial \Omega_t} \vec{g}(\vec{x},\vec{s})\, d \vec{x} = (0\,\,s_{49}lt\,\,0)^\trans$ along the top boundary $\partial \Omega_t$, where  $l \in \mathbb{R}^+$ is the total length of the cellular beam and $t \in \mathbb{R}^+$ is the thickness of the beam. The beam has a fixed length of $l=3000$ and fixed thickness of $t=10$ (both are deterministic). The (deterministic) height $b \in \mathbb{R}^+$ of the beam cross section is~$b=200$.

All random input variables $s_i$ are Gaussian, where $s_i \sim \mathcal{N}(\bar{s}_i,\sigma_i^2)$ and $\bar{s}_i \in \mathbb{R}$ is a constant for \mbox{$i = 49$}, and box-constrained $\bar{s}_i \in \{\bar{s}_i \in \mathbb{R}\, \vert \, \bar{s}_i^{(l)} \leq \bar{s}_i \leq \bar{s}_i^{(u)} \}$ for $i\in\{1,2,...,48\}$ where $\bar{s}_i^{(l)}$ and $\bar{s}_i^{(u)}$ are the lower and upper limits respectively. The lower and upper limits are defined using prior knowledge of the distribution of stress in the beam to ensure stress limits are not exceeded~\cite{bletzinger1990formoptimierung} (Table \ref{table:ex5_bounds}). Each design variable $s_i$ (where $i\in\{1,2,...,48\}$) is sampled from a Gaussian density with standard deviation $\sigma_i = 0.1$, and the uniformly distributed pressure load (immutable variable) is also random with a mean $\bar{s}_{49} = 3\times10^4$ and standard deviation $\sigma_{49}=5\times 10^3$.
\begin{table}[b!]
	\small
	\centering
	\caption{Cellular beam. Lower bounds $\bar{s}_i^{(l)}$ and upper bounds $\bar{s}_i^{(u)}$ for the $h$\textsuperscript{th}  void.}\label{table:ex5_bounds}
	\begin{tabular}{ccccccccc}
		\toprule
		$h$ & 1,16 & 2,15 & 3,14 & 4,13 & 5,12 & 6,11 & 7,10 & 8,9 \\
		\midrule
		$\bar{s}_{3h-2}^{(l)},\,\bar{s}_{3h-1}^{(l)}$ & 30 & 43 & 54 & 56 & 48 & 43 & 39 & 37 \\[0.6ex]
		$\bar{s}_{3h-2}^{(u)},\,\bar{s}_{3h-1}^{(u)}$ & 50 & 97 & 138 & 146 & 119 & 99 & 85 & 78 \\[0.6ex]
		$\bar{s}_{3h}^{(l)}$ & 5 & 9 & 13 & 14 & 11 & 9 & 8 & 7 \\[0.6ex]
		$\bar{s}_{3h}^{(u)}$ & 10 & 19 & 26 & 28 & 23 & 19 & 16 & 15 \\
		\bottomrule
	\end{tabular}
\end{table}

The goal is to determine the robust optimum design variables $\bar{\vec{s}}_d^* = (\bar{s}_1^*\,\,\bar{s}_2^*\,\,...\,\,\bar{s}_{48}^*)^\trans$ that minimise structural compliance \eqref{eq:e10}, subject to a volume constraint \eqref{eq:e11} with the limit volume $V_0 = 5.2 \times 10^{6}$. The weighting factor and the feasibility index are chosen as $\alpha=0.25$ and $\beta^{(1)}=2$.
The training data set $\mathcal{D}$ is initialised using LHS with $n=3000$ samples, and $m=300$ inducing points. The RDVGP surrogate is trained using $n_r=20$ restarts and $n_t = 10^4$ iterations using stochastic gradient descent (Algorithm \ref{alg:rdvgp}). The compliance  $J(\vec{s})$  is computed by solving the elasticity equations using the FE model. The cellular beam is modelled as a thin plate using the plane stress assumption. The Young's modulus and Poisson's ratio are $2.1\times10^{5}$  and  $0.3$. The normalisation constants $\bar{\mu}$ and $\bar{\sigma}$ from the RDO formulation are set equal to the mean and standard deviation of observations of the objective function $J(\vec{s})$. A linear subspace consisting of ten dimensions ($d_z=10$) is chosen. Although the dimension may be insufficient to fully describe the variance in the compliance $J(\vec{s})$ and volume constraint $H^{(1)}(\vec{s})$, it provides an approximation that captures most of the variance without compromising computational efficiency.

\begin{figure*}[b!]
		\centering
		\subfloat[Stress mean $\expect\left(\sigma_v(\vec{x},\vec{s}^*)\right)$]{\includegraphics[width=110mm]{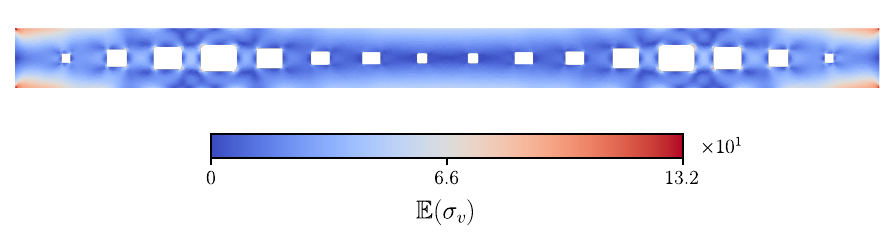}}\\
		\centering
		\subfloat[Stress standard deviation $\sqrt{\var\left(\sigma_v(\vec{x},\vec{s}^*)\right)}$]{\includegraphics[width=110mm]{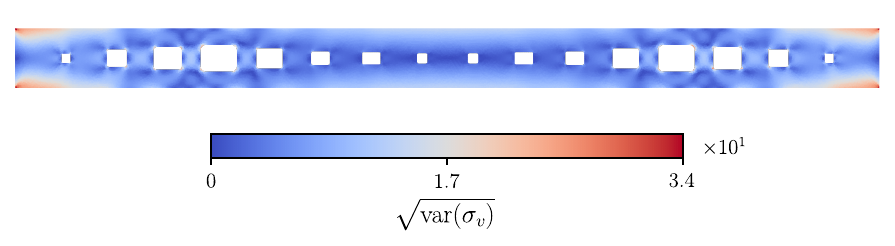}}
		\caption{Cellular beam. Deformation showing for the optimum geometry $\vec{s}^* = ((\bar{\vec{s}}_d^*)^\trans\,\,s_f)^\trans$ (a) the von Mises stress mean $\expect\left(\sigma_v(\vec{x},\vec{s}^*)\right)$, (b) the von Mises stress standard deviation $\sqrt{\var\left(\sigma_v(\vec{x},\vec{s}^*)\right)}$, (c) the displacement mean $\expect\left(\vert\vert\vec{u}(\vec{x},\vec{s}^*)\vert\vert_2\right)$, and (d) the displacement standard deviation $\sqrt{\var\left(\vert\vert\vec{u}(\vec{x},\vec{s}^*)\vert\vert_2\right)}$.}
		\label{fig:ex5_res}
\end{figure*}

A robust minimum is obtained by performing RDO with the trained RDVGP surrogate, which shows the uniformly distributed pressure load $s_{49}$ and the height of the voids $s_{3h-1}$ are the input variables that most influence the compliance $J(\vec{s})$. The void fillet radii $s_{3h}$ have negligible influence on either the compliance or volume $H^{(1)}(\vec{s})$ of the beam. Since both the width $s_{3h-2}$ and height $s_{3h-1}$ of the voids have the same effect on volume but different effects on compliance, the robust optimum solution shows the voids are greater in width than they are in height (Figure \ref{fig:ex5_res}). Compared to the nominal solution (which is computed by scaling each of the design parameters uniformly between their upper and lower limits until the limit volume is obtained), the maximum mean displacement is reduced from 1.76 to 1.65 and the standard deviation reduced from 0.48 to 0.44. This shows that RDO with the RDVGP surrogate can improve stiffness without increasing the volume of material.

%% file: conclusion.tex
\section{Conclusion}\label{section:conclusion}

We introduced the RDVGP surrogate, which provides a quick-to-evaluate approximation of the outputs of complex computational models. The construction of the surrogate is formulated as a Bayesian inference problem. The assumed statistical observation model includes a low-dimensional latent vector that provides an inherent dimensionality reduction. The conditional independence structure of the RDVGP surrogate can be visualised as a directed graphical model. Unlike prevalent surrogate modelling approaches, the inputs to the RDVGP surrogate are random and have a prescribed covariance. The RDVGP surrogate is trained by fitting to data from a deterministic black-box computational model. The respective posterior probability density represents the surrogate and is approximated by a trial Gaussian density with a diagonal covariance structure using variational Bayes. Both the statistical model hyperparameters and the trial density variational parameters are determined by maximising the ELBO. The hyperparameters of the RDVGP model include the entries of the orthogonal projection matrix mapping the inputs to the latent vector, and the covariance parameters of the prior density. Overall, the computational simplicity of the RDVGP model provides a favourable alternative to common surrogate modelling approaches, which usually do not consider input uncertainty. Unlike for instance standard GP regression, the RDVGP surrogate avoids over-fitting and yields a principled representation of epistemic model uncertainty and aleatoric output uncertainty. Although we demonstrated the performance of the RDVGP surrogate only for RDO applications, it is equally suitable for other multi-query applications involving random inputs.

In closing, we highlight several potential generalisations and extensions to the introduced surrogate modelling approach. In RDO the input covariance is prescribed, but in other applications the input covariance may be unknown. In such cases, the input covariance can be estimated by treating it as an additional model hyperparameter and maximising the ELBO using stochastic gradient descent.   It is usually beneficial to consider data from computational models with varying fidelity, e.g. with  different discretisations, to improve training efficiency of surrogates. The additive noise (or error) component in the presented statistical observation model can take into account the associated modelling errors. The training efficiency and expressivity of the RVDGP surrogate can be further increased by replacing the posited GP prior with more advanced multifideliy GP or deep GP priors~\cite{kennedy2000predicting, damianou2013deep}.  Alternatively, it may be expedient to consider standard GP priors with a mean and covariance parameterised by a neural network~\cite{rixner2021probabilistic, vadeboncoeur2023fully}. Furthermore, considering non-linear dimensionality reduction techniques, see, e.g. \cite{lee2007nonlinear}, may reduce the number of latent variables, improving efficiency and accuracy for specific problems. Finally, the use of the RDVGP surrogate in combination with adaptive sampling algorithms, such as in Bayesian optimisation or active learning approaches~\cite{forrester2008engineering,sauer2023active}, represents a promising direction for further research.

%% file: appendix.tex
\hypertarget{Appendix A}{\section*{Appendix A : Derivation of the KL divergence terms}\label{section:appendix_kl_deriv}}

The KL divergence terms in the ELBO $\mathcal{F}(\vec{y})$ given by \eqref{eq:splvm_8} are analytically tractable and can be computed in exact form. For example, consider the second KL divergence term 
\begin{equation*}\label{eq:kl1}
\scalebox{0.7}{$
\begin{aligned}
D_{KL}\left( q_{\psi}(\vec{Z}) \,\vert\vert\, p_{W}(\vec{Z}) \right) &= \int q_{\psi}(\vec{Z}) \ln \left(\frac{q_{\psi}(\vec{Z})}{p_{W}(\vec{Z})}\right)d\vec{Z}\\
&= \sum_{i=1}^{n}\expect_{q_{\psi}(z_i)}\left( \ln q_{\psi}(\vec{z}_i) - \ln p_{W}(\vec{z}_i) \right)\\
&= \frac{1}{2}\sum_{i=1}^{n}\expect_{q_{\psi}(z_i)}\left( \ln \left(\frac{\vert \vec{W}^\trans\vec{\Sigma}_s\vec{W} \vert}{\vert \tilde{\vec{\Sigma}}_{z,i} \vert}\right) + \left(\vec{z}_i-\vec{W}^\trans\bar{\vec{s}}_i\right)^\trans \left(\vec{W}^\trans\vec{\Sigma}_s\vec{W}\right)^{-1}\left(\vec{z}_i-\vec{W}^\trans\bar{\vec{s}}_i\right) - \left(\vec{z}_i-\tilde{\vec{\mu}}_{z,i}\right)^\trans \left(\tilde{\vec{\Sigma}}_{z,i}\right)^{-1}\left(\vec{z}_i-\tilde{\vec{\mu}}_{z,i}\right)\right)\\
&= \frac{1}{2}\sum_{i=1}^{n}\left( \ln \left(\frac{\vert \vec{W}^\trans\vec{\Sigma}_s\vec{W} \vert}{\vert \tilde{\vec{\Sigma}}_{z,i} \vert}\right) + \expect_{q_{\psi}(z_i)}\left( 
\trace\left( \left(\vec{W}^\trans\vec{\Sigma}_s\vec{W}\right)^{-1}\left(\vec{z}_i-\vec{W}^\trans\bar{\vec{s}}_i\right)\left(\vec{z}_i-\vec{W}^\trans\bar{\vec{s}}_i\right)^\trans\right) - \trace\left( \left(\tilde{\vec{\Sigma}}_{z,i}\right)^{-1}\left(\vec{z}_i-\tilde{\vec{\mu}}_{z,i}\right)\left(\vec{z}_i-\tilde{\vec{\mu}}_{z,i}\right)^\trans\right) \right)\right)\\
&= \frac{1}{2}\sum_{i=1}^{n}\left( \ln \left(\frac{\vert \vec{W}^\trans\vec{\Sigma}_s\vec{W} \vert}{\vert \tilde{\vec{\Sigma}}_{z,i} \vert}\right) + \expect_{q_{\psi}(z_i)}\left( 
\trace\left( \left(\vec{W}^\trans\vec{\Sigma}_s\vec{W}\right)^{-1}\left(\vec{z}_i\vec{z}_i^\trans-2\vec{z}_i\bar{\vec{s}}_i^\trans\vec{W} + \vec{W}^\trans\bar{\vec{s}}_i\bar{\vec{s}}_i^\trans\vec{W}\right)\right) - \trace\left( \left(\tilde{\vec{\Sigma}}_{z,i}\right)^{-1}\tilde{\vec{\Sigma}}_{z,i}\right) \right)\right)\\
&= \frac{1}{2}\sum_{i=1}^{n}\left( \ln \left(\frac{\vert \vec{W}^\trans\vec{\Sigma}_s\vec{W} \vert}{\vert \tilde{\vec{\Sigma}}_{z,i} \vert}\right) - d_z +  
\trace\left( \left(\vec{W}^\trans\vec{\Sigma}_s\vec{W}\right)^{-1}\left(\tilde{\vec{\Sigma}}_{z,i} + \tilde{\vec{\mu}}_{z,i}\left(\tilde{\vec{\mu}}_{z,i}\right)^\trans - 2\vec{W}^\trans\bar{\vec{s}}_i\left(\tilde{\vec{\mu}}_{z,i}\right)^\trans + \vec{W}^\trans\bar{\vec{s}}_i\bar{\vec{s}}_i^\trans\vec{W}\right)\right) \right)\\
&= \frac{1}{2}\sum_{i=1}^{n}\left( \ln \left(\frac{\vert \vec{W}^\trans\vec{\Sigma}_s\vec{W} \vert}{\vert \tilde{\vec{\Sigma}}_{z,i} \vert}\right) - d_z + \trace\left(\left(\vec{W}^\trans\vec{\Sigma}_s\vec{W}\right)^{-1}\tilde{\vec{\Sigma}}_{z,i}\right)  + \left(\vec{W}^\trans\bar{\vec{s}}_i-\tilde{\vec{\mu}}_{z,i}\right)^\trans\left(\vec{W}^\trans\vec{\Sigma}_s\vec{W}\right)^{-1}\left(\vec{W}^\trans\bar{\vec{s}}_i-\tilde{\vec{\mu}}_{z,i}\right)
 \right),
\end{aligned}
$}\tag{A.1}
\end{equation*}
and after following a similar derivation for the first KL divergence term
\begin{equation*}\label{eq:kl2}
\begin{aligned}
D_{KL}\left( q_{\omega}(\tilde{\vec{f}}) \,\vert\vert\, p_{\theta}(\tilde{\vec{f}}) \right) 
&= \frac{1}{2}\left( \ln \left(\frac{\vert \vec{C}_{\tilde{Z}\tilde{Z}} \vert}{\vert \tilde{\vec{\Sigma}}_{\tilde{f}} \vert}\right) - m + \trace\left(\vec{C}_{\tilde{Z}\tilde{Z}}^{-1} \, \tilde{\vec{\Sigma}}_{\tilde{f}}\right)  + (\tilde{\vec{\mu}}_{\tilde{f}})^\trans\vec{C}_{\tilde{Z}\tilde{Z}}^{-1} \,\tilde{\vec{\mu}}_{\tilde{f}}
 \right).
\end{aligned}
\tag{A.2}
\end{equation*}

\hypertarget{Appendix B}{\section*{Appendix B: Multiple observation vectors}\label{section:appendix_multiple}}

The $d_y$ observation vectors and corresponding target output variables can be collected into the matrices $\vec{Y} \in \mathbb{R}^{n \times d_y}$ and $\vec{F} \in \mathbb{R}^{n \times d_y}$ respectively, where $\vec{Y}=\left(\vec{y}^{(1)}\,\,\vec{y}^{(2)}\,\,...\,\,\vec{y}^{(d_y)}\right)$ and $\vec{F}=\left(\vec{f}^{(1)}\,\,\vec{f}^{(2)}\,\,...\,\,\vec{f}^{(d_y)}\right)$. The likelihood
\begin{equation*}\label{eq:mo1}
\begin{aligned}
p_{\sigma_y}(\vec{Y}\vert\vec{F}) = \prod_{i=1}^{d_y} p_{\sigma_y}\left(\vec{y}^{(i)}\big\vert\vec{f}^{(i)}\right)=\prod_{i=1}^{d_y} \mathcal{N}\left(\vec{f}^{(i)},\sigma_{y}^{(i)}\vec{I}\right),
\end{aligned}
\tag{B.1}
\end{equation*}
and GP prior probability density

\begin{equation*}\label{eq:mo2}
\begin{aligned}
p_\theta(\vec{F}\vert\vec{Z}) = \prod_{i=1}^{d_y} p_{\theta}\left(\vec{f}^{(i)}\big\vert\vec{Z}\right)= \prod_{i=1}^{d_y} \mathcal{N}\left(\vec{0},\vec{C}_{ZZ}^{(i)}\right),
\end{aligned}
\tag{B.2}
\end{equation*}
can be expanded using the mean field approximation. Similarly, the ELBO $\mathcal{F}(\vec{y})$ in \eqref{eq:plvm_13} for the statistical latent variable model may be expanded as the summation of the ELBO over each observation vector
\begin{equation*}\label{eq:mo3}
\begin{aligned}
\mathcal{F}(\vec{y})
&= \sum_{i=1}^{d_y}\expect_{q_\psi(\vec{Z})}\Bigg( \expect_{p_\theta(\vec{f}^{(i)}\vert\vec{Z})}\left( \ln p_{\sigma_y}\left(\vec{y}^{(i)} \big\vert \vec{f}^{(i)}\right) \right) \Bigg) - D_{KL}\left(q_{\psi}(\vec{Z}) \vert \vert p_{W}(\vec{Z})\right).
\end{aligned}
\tag{B.3}
\end{equation*}
Similarly for the sparse statistical latent variable model, the sparse GP prior 
\begin{equation*}\label{eq:mo4}
\begin{aligned}
p_\theta\left(\vec{F}\vert\tilde{\vec{F}},\vec{Z}\right) = \prod_{i=1}^{d_y} p_{\theta}\left(\vec{f}^{(i)}\vert \tilde{\vec{f}}^{(i)} ,\vec{Z}\right) = \prod_{i=1}^{d_y} \mathcal{N}\left(\vec{C}_{Z\tilde{Z}}^{(i)}\left(\vec{C}_{\tilde{Z}\tilde{Z}}^{(i)}\right)^{-1}\tilde{\vec{f}}^{(i)}\,,\, \vec{C}_{ZZ}^{(i)}-\vec{C}_{Z\tilde{Z}}^{(i)}\left(\vec{C}_{\tilde{Z}\tilde{Z}}^{(i)}\right)^{-1}\vec{C}_{\tilde{Z}Z}^{(i)}\right),
\end{aligned}
\tag{B.4}
\end{equation*}
and trial probability density
\begin{equation*}\label{eq:mo5}
\begin{aligned}
q_\omega\left(\tilde{\vec{F}}\right) := \prod_{i=1}^{d_y}q_{\omega}\left(\tilde{\vec{f}}^{(i)}\right) := \prod_{i=1}^{d_y}\mathcal{N}\left(\tilde{\vec{\mu}}_{\tilde{f}}^{(i)},\tilde{\vec{\Sigma}}_{\tilde{f}}^{(i)}\right),
\end{aligned}
\tag{B.5}
\end{equation*}
can be expanded, where $\tilde{\vec{F}} \in \mathbb{R}^{m \times d_y}$ is the matrix of pseudo target output variables.

\hypertarget{Appendix C}{\section*{Appendix C : Comparison of RDVGP and GP surrogates}\label{section:rdvgp_gp_comparison}}

We compare the standard GP and RDVGP surrogates in the one-dimensional illustrative example (section \ref{subsection:illustrative_example_1}). The standard GP surrogate fitted using two advanced python libraries GPyTorch~\cite{gardner2018gpytorch} and GP+~\cite{yousefpour2024gp+}, is compared to the RDVGP surrogate introduced in this paper. The surrogates are compared across training data consisting of $n=5$, $n=8$, and $n=11$ samples. The RDVGP surrogate is fitted with $m=n$ inducing points.

\begin{figure*}[b!]
	\centering
	\subfloat[GPyTorch]{\includegraphics[width=162mm]{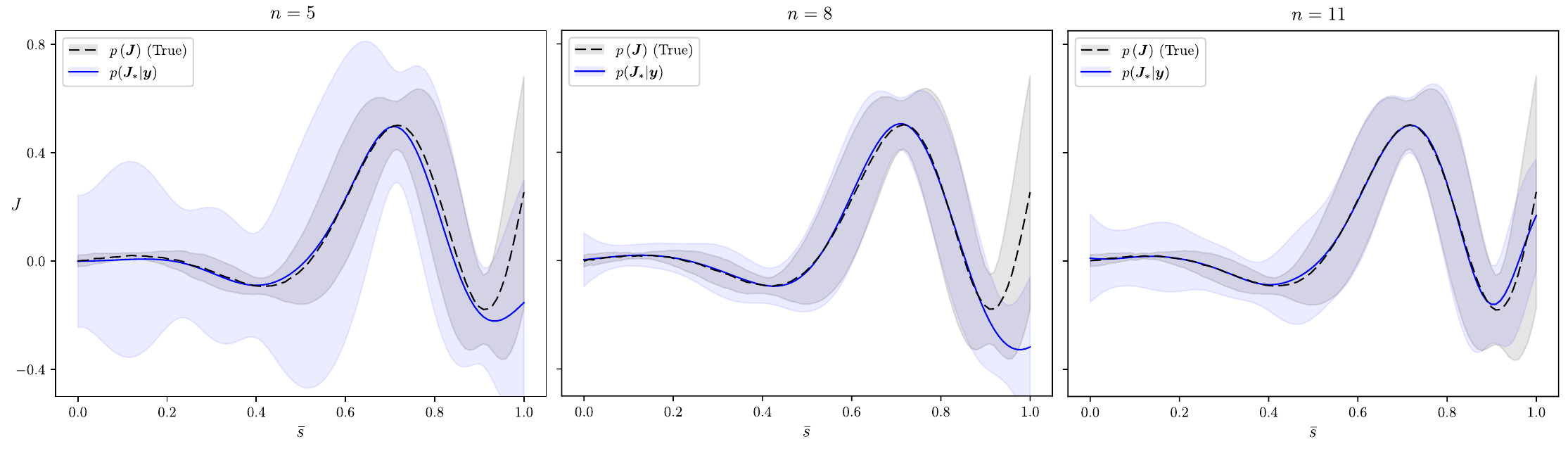}}\\
	\centering
	\subfloat[GP+]{\includegraphics[width=162mm]{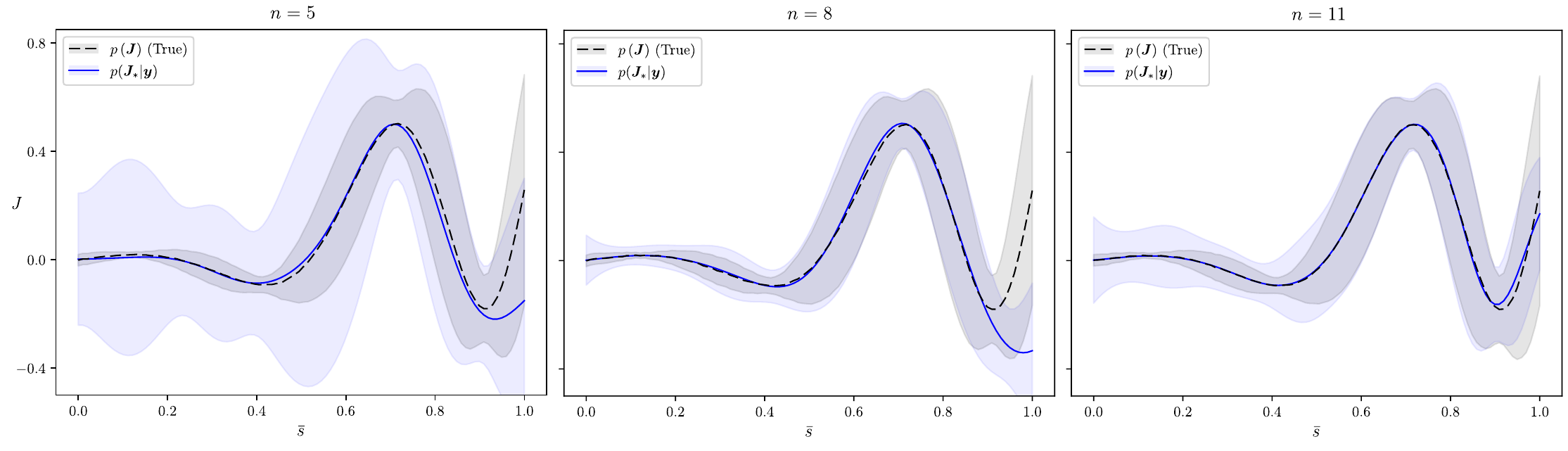}}\\
	\centering
	\subfloat[RDVGP]{\includegraphics[width=162mm]{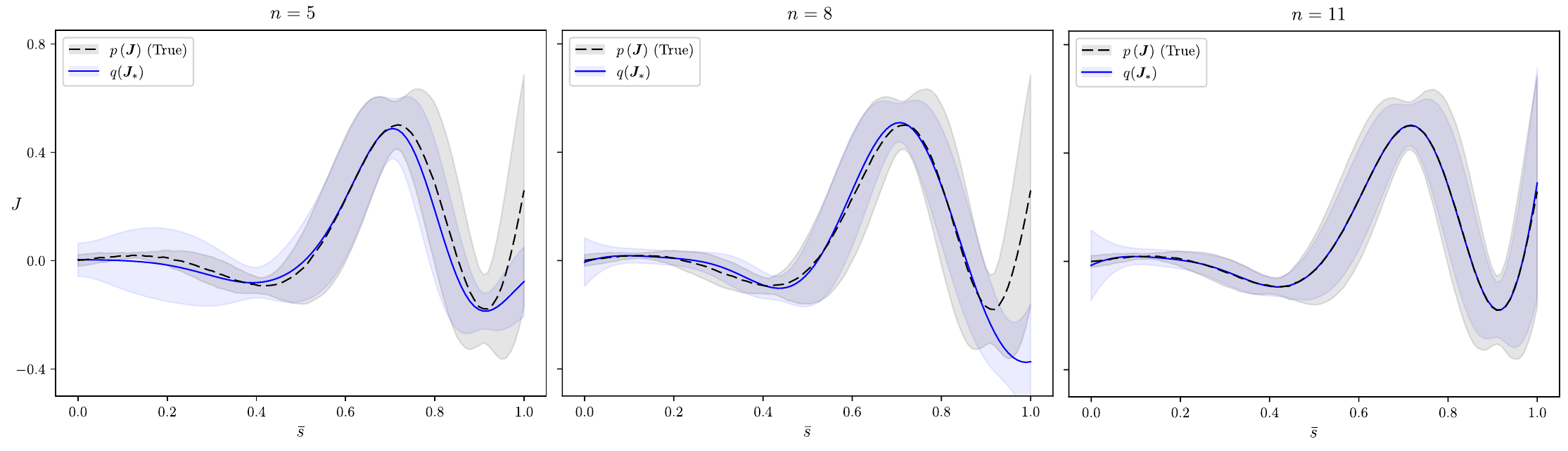}}
	\caption{Surrogate comparison. Comparison of the standard GP surrogates fitted using (a) GPyTorch, (b) GP+, and (c) with the RDVGP surrogate, using $n=5$, $n=8$, and $n=11$ samples.}
	\label{fig:surrogate_comp}
\end{figure*}

With a small data set (i.e. $n=5$), the standard GP both overfits the training data, as observed with both the GPyTorch and GP+ libraries, see e.g. Figure \ref{fig:surrogate_comp}. However, with a larger data set (i.e. $n=8$ and $n=11$), the difference between the RDVGP and standard GP surrogates decreases, although the RDVGP surrogate still maintains a better fit. The posterior density $p(\vec{J}_*\,\vert\,\vec{y})$ obtained using the GPyTorch library is near-identical to the posterior obtained using the GP+ library, although the GP+ library obtains a more competitive performance for advanced problems~\cite{yousefpour2024gp+}.

